%% file: main.tex
\PassOptionsToPackage{pdfversion=2.0,pdfa}{hyperref}
\PassOptionsToPackage{hyphens}{url} 
\documentclass[sigconf]{acmart}
\hypersetup{pdfapart=1,pdfaconformance=B}
\input{macros}
\copyrightyear{2026}
\acmYear{2026}
\setcopyright{cc}
\setcctype{by}
\acmConference[WSDM '26]{Proceedings of the Nineteenth ACM International Conference on Web Search and Data Mining}{February 22--26, 2026}{Boise, ID, USA}
\acmBooktitle{Proceedings of the Nineteenth ACM International Conference on Web Search and Data Mining (WSDM '26), February 22--26, 2026, Boise, ID, USA}
\acmPrice{}
\acmDOI{10.1145/3773966.3777942}
\acmISBN{979-8-4007-2292-9/2026/02}

\title{\rwcpp: A Statistically-Principled Structural Polarization Measure}

\author{Giulia Preti}
 \orcid{0000-0002-2126-326X} 
\affiliation{%
  \institution{CENTAI}
  \city{Turin}
  \country{Italy}
}
\email{giulia.preti@centai.eu}

\author{Matteo Riondato}
 \orcid{0000-0003-2523-4420} 
\affiliation{%
  \institution{Amherst College}
  \city{Amherst}
  \state{MA}
  \country{USA}}
\email{mriondato@amherst.edu}

\author{Aristides Gionis}
 \orcid{0000-0002-5211-112X} 
\affiliation{%
  \institution{KTH Royal Institute of Technology}
  \city{Stockholm}
  \country{Sweden}}
\email{argioni@kth.se}

\author{\texorpdfstring{\nohyphens{Gianmarco De~Francisci Morales}}{Gianmarco De Francisci Morales}}
 \orcid{0000-0002-2415-494X} 
\affiliation{%
 \institution{CENTAI}
 \city{Turin}
 \country{Italy}}
\email{gdfm@acm.org}

\begin{document}

\begin{abstract}
  Social and information networks may become polarized, leading to echo chambers
  and political gridlock. Accurately measuring this phenomenon is a critical
  challenge. Existing measures often conflate genuine structural division with
  random topological features, yielding misleadingly high polarization scores on
  random networks, and failing to distinguish real-world networks from randomized
  null models. We introduce \rwcpp, a Diffusion-based Structural Polarization
  measure designed from first principles to correct for such biases.
  \rwcpp removes the arbitrary concept of ``influencers'' used by the
  popular Random Walk Controversy (\rwc) score, instead treating every node as
  a potential origin for a random walk. To validate our approach, we introduce a
  set of desirable properties for polarization measures, expressed through
  reference topologies with known structural properties. We show that \rwcpp
  satisfies these desiderata, being near-zero for non-polarized structures such
  as cliques and random networks, while correctly capturing the expected
  polarization of reference topologies such as monochromatic-splittable
  networks. Our method applied to U.S.~Congress datasets uncovers trends of increasing polarization in recent years.
  By integrating a null model into its core definition,
  \rwcpp provides a reliable and interpretable diagnostic tool, highlighting the necessity of statistically-grounded metrics to analyze societal fragmentation.
\end{abstract}



\maketitle

\input{sections/intro}
\input{sections/prelims}

\input{sections/algo}
\input{sections/analysis}
\input{sections/exp}
\input{sections/conclusion}


\begin{acks}
  MR's work is supported by the \grantsponsor{NSF}{National Science
  Foundation}{https://www.nsf.gov} award
  CAREER-\grantnum[https://www.nsf.gov/awardsearch/showAward?AWD_ID=2238693]{NSF}{2238693}.
  AG's work is supported by the ERC Advanced Grant REBOUND [834862],  the Swedish Research Council (VR) [2024-05603], the European Commission MSCA DN [101168951], and the Wallenberg AI, Autonomous Systems and Software Program (WASP) funded by the Knut and Alice Wallenberg Foundation.
\end{acks}

\clearpage
\input{sections/ethics}

\bibliographystyle{ACM-Reference-Format}
\balance%
\bibliography{biblio}

 \appendix

 \input{sections/appendix}

\end{document}

%% file: macros.tex
\usepackage{graphicx} 
\usepackage{xspace} 
\usepackage{multirow} 
\usepackage[font={bf}, tableposition=top]{caption} 
\usepackage{subcaption} 
\usepackage{bold-extra} 
\usepackage[vlined,linesnumbered,ruled,noend]{algorithm2e} 
\usepackage{microtype} 
\usepackage{siunitx} 
\usepackage{xfrac} 
\usepackage{mathtools} 
\usepackage{amsthm}
\usepackage{enumitem}
\usepackage[bb=dsserif]{mathalpha} 
\usepackage{bm} 
\hypersetup{colorlinks=true}
\usepackage{cleveref} 
\usepackage{hyphenat} 
\usepackage[show]{chato-notes} 

\newcommand{\spara}[1]{\smallskip\noindent\textbf{#1}}
\newcommand{\epara}[1]{\noindent\emph{#1}}

\newenvironment{squishlist}
{\begin{list}{$\bullet$}
 {\setlength{\itemsep}{0pt}
     \setlength{\parsep}{3pt}
     \setlength{\topsep}{3pt}
     \setlength{\partopsep}{0pt}
     \setlength{\leftmargin}{1.5em}
     \setlength{\labelwidth}{1em}
     \setlength{\labelsep}{0.5em} } }
{\end{list}}

\usepackage{xcolor}

\graphicspath{{figures/}{img/}}

\DeclarePairedDelimiter\abs{\lvert}{\rvert} 
\makeatletter
\let\oldabs\abs%
\def\abs{\@ifstar{\oldabs}{\oldabs*}}
\makeatother

\newcommand{\er}{Erd\H{o}s-R\'{e}nyi\xspace}

\newcommand{\arwc}{\ensuremath{\text{ARWC}}\xspace}
\newcommand{\gnml}{\ensuremath{G(n,p,\mathbf{\ell})}\xspace}

\newcommand{\rwc}{\ensuremath{\text{RWC}}\xspace}
\newcommand{\rwcpp}{\ensuremath{\text{DSP}}\xspace}
\newcommand{\sizeof}[1]{\ensuremath{\abs{#1}}\xspace}

\DeclarePairedDelimiter\set{\{}{\}} 
\makeatletter
\let\oldset\set%
\def\set{\@ifstar{\oldset}{\oldset*}}
\makeatother
\newcommand{\procfun}[1]{\ensuremath{\phi_{#1}}\xspace}
\newcommand{\procprobfun}[1]{\ensuremath{\pi_{#1}}\xspace}
\newcommand{\procprob}[2]{\ensuremath{\procprobfun{#1}(#2)}\xspace} 
\newcommand{\community}{\ensuremath{\mathbf{Q}}\xspace}
\newcommand{\source}{\ensuremath{\mathbf{S}}\xspace}
\newcommand{\target}{\ensuremath{\mathbf{T}}\xspace}
\newcommand{\probe}{\ensuremath{\mathbf{Y}}\xspace}
\newcommand{\reds}{\ensuremath{R}\xspace}
\newcommand{\blues}{\ensuremath{B}\xspace}
\newcommand{\col}[1]{\ensuremath{\mathsf{c}(#1)}\xspace}

%% file: sections/intro.tex
\section{Introduction}
\label{sec:intro}

Quantifying the structural polarization of a network from its topology is crucial for identifying systemic risks---from political gridlock~\citep{poole2001d} to the proliferation of misinformation~\citep{nikolov2021right}---and for developing effective interventions to mitigate these harmful effects~\citep{matakos2017measuring}.
This phenomenon is characterized by the grouping of individuals into coherent communities that rarely interact with each other~\citep{conover2011political}, and it is increasingly observed across social, political, and information networks, where its effects can be seen in the emergence of ideological echo chambers~\citep{cinelli2021echo} and adversarial dynamics~\citep{cossard2020falling}.
Mitigating the detrimental effects of polarization on collective decision-making 
and overall societal stability necessitate robust methods for its quantification and analysis~\citep{barber2015causes}.

Existing measures of structural polarization, however, face significant limitations.
Many traditional methods, such as modularity or assortativity, rely on simplistic null models that often fail to capture the complex dynamics of the real-world networks~\citep{guerra2013measure,leifeld2018polarization}.
Other more recent measures, such as Betweenness Centrality Controversy~\citep{garimella2018quantifying}, Boundary Polarization~\citep{guerra2013measure}, Dipole Polarization~\citep{morales2015measuring}, Krackhardt E/I Ratio~\citep{krackhardt1988informal}, and Random Walk Controversy~\citep{garimella2018quantifying}, can produce high polarization scores even for random networks, indicating an undesirable sensitivity to elementary network features such as average degree and network size~\citep{salloum2022separating}.
These biases occur because existing controversy measures conflate topological artifacts, such as low density, small network size, or uneven group splits, with genuine structural division.


Attempts to correct these biases, such as post-hoc normalization via randomization tests~\citep{salloum2022separating}, remain fundamentally unprincipled, as they apply arbitrary null models retroactively without rigorous justification.
As such, they risk over- or under-correction, as different null assumptions (e.g., preserving degree sequences vs. erasing all structure) yield conflicting baselines~\citep{preti2022alice}.
Moreover, normalization is treated as an external adjustment rather than a core component of the measure's design, perpetuating ambiguities in score interpretation.

\begin{table*}[t]
    \centering
    \caption{Most common structural polarization measures.}\label{tab:pol_measures}
    \resizebox{\linewidth}{!}{
        \begin{tabular}{lll}
        \toprule
        \textbf{Measure} & \textbf{Range} & \textbf{Intuition} \\
        \midrule
        Random Walk Controversy (\rwc)~\citep{garimella2018quantifying} & $[-1, 1]$ & In polarized networks, users are less exposed to cross-cutting content.\\
        Adaptive Random Walk Controversy (ARWC)~\citep{garimella2018quantifying} & $[-1, 1]$ & RWC, adjusting for community size in the number of influencers.\\
        Betweenness Centrality Controversy (BCC)~\citep{garimella2018quantifying} & $[0, 1]$ & High betweenness centrality on boundary edges indicates separation.\\
        Boundary Polarization (BP)~\citep{guerra2013measure} & $[-0.5, 0.5]$ & In polarized networks, authoritative users are further from the boundary.\\
        Cross-community Affinity (CCA)~\citep{nair2024cross} & $[-1.5, 1.5]$ & Direct and indirect links influence a node's ideological openness and cross-community affinity.\\
        Color Assortativity (Col-Ass)~\citep{newman2003mixing} & $[-1, 1]$ & Higher tendency to connect with nodes with the same opinion indicates separation.\\
        Dipole Moment (DM)~\citep{morales2015measuring} & $[0, 1]$ & Greater distance between positive and negative opinions indicates separation.\\
        Krackhardt E/I Ratio (EI)~\citep{krackhardt1988informal} & $[-1, 1]$ & Higher fraction of within-community edges indicates separation.\\ 
        Adaptive E/I Index (AEI)~\citep{chen2021polarization} & $[-1, 1]$ & EI, accounting for different community sizes.\\
        Modularity (Q)~\citep{waugh2009party} & $[-0.5, 1]$ & More within-community edges than expected by chance indicates separation.\\
        \bottomrule
        \end{tabular}
    }
\end{table*}

To address the aforementioned issues, we introduce \rwcpp, a new measure for structural polarization, built on a principled statistical model.
Unlike previous proposals, \rwcpp is designed to have an expected value of zero on \gnml networks, a labeled extension of the \er \ensuremath{G(n,p)} model where edges and labels are assigned randomly, thus ensuring no structural correlation exists between the two.
Our design explicitly removes the biases found in the original Random Walk Controversy (\rwc)~\citep{garimella2018quantifying}, by eliminating the problematic role of predefined influencers.
Instead, \rwcpp treats each vertex as a potential source for a random walk, with all other vertices as targets.
This approach retains the strength of \rwc in being independent of partition sizes and in leveraging random walks to measure distance and information spread effectively.
Nonetheless, by considering every vertex a potential source, \rwcpp becomes more robust to different network structures and better reflects how information spreads across the network (see \Cref{sec:analysis}).

Eliminating the idea of designating a small number of vertices as influencers and considering all vertices as potential influencers poses new challenges, as a random walk (re)starting from a given vertex $v$ has high probability of visiting vertex $v$ more frequently, which introduces a new source of bias. 
We show how to avoid this bias by designing a probing process tailored to our task. 
The resulting \rwcpp measure can be interpreted as the average score of how likely
a vertex is to receive information from a given community, 
where the average is taken over the distribution defined by our probing process.
As a result, \rwcpp is statistically more principled than \rwc, 
which incorporates products of probabilities with no clear statistical interpretation.

We validate our approach through extensive experiments on both synthetic and real-world networks.
First, we develop a set of reference networks with prescribed values of structural polarization and confirm that \rwcpp behaves as expected.
On \gnml networks, \rwcpp exhibits near-zero scores as theoretically guaranteed.
Second, for real-world networks, \rwcpp distinguishes ideologically polarized communities from structurally similar but neutralized configurations, further demonstrating its robustness to group size imbalances and addressing limitations of both \rwc and traditional metrics.
Furthermore, we analyze the relationship between assortativity and polarization as captured by \rwcpp. Finally, we show that \rwcpp can be approximated efficiently with good accuracy.

By integrating a statistically principled null model into its core formulation, \rwcpp advances polarization measurement beyond ad hoc corrections.
This approach answers a critical need for metrics grounded in explicit, theoretically sound baselines, which is essential for developing reliable diagnostics in an era of algorithmic fragmentation and polarized discourse.
Our findings bridge key gaps in existing methodologies, enriching the toolkit for analyzing structural polarization and highlighting the value of robust measures to guide effective interventions in complex social landscapes.


%% file: sections/prelims.tex
\section{Background and Preliminaries}\label{sec:prelims}

Structural polarization measures aim at quantifying whether a given network represents a polarized system from its topology.
Because polarization is a system-level phenomenon, these features are typically defined at the network level rather than the vertex level (such as individual behavioral mechanisms).
Similar measures have been defined for economic systems, e.g., the Gini coefficient and the Theil index, as well as in other domains (e.g., spatial, occupational, and educational segregation); however, these measures deal with numerical data rather than networks.
\Cref{tab:pol_measures} reports the structural polarization measures most commonly used to analyse social networks, together with an intuition of the measure's aim.

\spara{Structural polarization pipeline}.
Structural polarization measures are designed to be used as part of a broader pipeline~\citep{garimella2016quantifying,salloum2022separating}.
The typical pipeline comprises the following three stages:
\begin{squishlist}
\item[($i$)] Define an appropriate directed network, e.g., a web graph, a friendship network, or an endorsement network.
\item[($ii$)] Partition this network into (typically two) separate communities, under the assumption that the network reflects polarized opinions along a single ideological dimension;
\item[($iii$)] Compute the structural polarization measure starting from the communities from the previous step.
\end{squishlist}
The focus of the current work is on the third and last step.
While this pipeline is commonly used, it is by no means the only way to use structural polarization measures.
For instance, communities might already be defined in the data or derived from distant labels (e.g., semantically polarized hashtags).

\begin{table}[t]
\caption{Summary of notation.}%
\label{tab:notation}
\centering
\small
\begin{tabular}{lp{6.5cm}}
\toprule
\text{Symbol} & Meaning \\
\midrule
$V$ & Set of vertices in the network\\
$E$ & Set of edges in the network ($E \subseteq V \times V)$\\
$A$ & Network adjacency matrix \\ 
$n$ & Number of vertices ($n = \sizeof{V}$) \\
\reds, \blues & Partitions of V ($\reds \cup \blues=V, \reds \cap \blues = \varnothing$)\\
\col{v} & Partition of vertex $v$, referred to as its \emph{color} ($\col{v} \in \left\{\reds,\blues\right\}$)\\
\procfun{Z} & Stationary distribution of a diffusion process rooted in $Z \subseteq V$ (e.g., Random Walk with Restart)\\
\bottomrule
\end{tabular}
\end{table}

\subsection{Problem Setting and Notation}\label{sec:notation}

This section introduces our notation (summarized in \Cref{tab:notation}).
We consider a weakly-connected directed network $G=(V,E)$, 
where $n=\sizeof{V}$.
The set of vertices $V$ is partitioned into two subsets, \reds and \blues, 
either algorithm\-ical\-ly (e.g., by applying a graph-partitioning algorithm) or by metadata (e.g., by using vertex features).
We refer to these sets as \emph{communities}.
We call \emph{vertex color} the attribute of a vertex $v \in V$ that indicates membership to a given community,
denoted as $\col{v} \in \left\{\reds,\blues\right\}$.
$A$ denotes the adjacency matrix of the network $G$, 
i.e., $A_{ij} = 1$ if $(i,j) \in E$ and $A_{ij} = 0$ otherwise. 
If the edges of $G$ have weights, they can be incorporated in the corresponding entries of the adjacency matrix $A$.

We assume a function $\procfun{v}: V \to \mathbb{R}^+$ 
that assigns a non-negative value to each vertex $w\in V$, parameterized by a vertex
$v \in V$. The function $\procfun{v}$ is arbitrary, provided it does not depend
on the colors of the vertices. Intuitively, it captures aspects of the network
structure, so that more important vertices $w$ tend to have higher values.
Specifically, it represents a network diffusion process~\citep{masuda2017random} 
and measures how information originating from $v$ spreads.

As a concrete example, we consider the stationary distribution of the random
walk with restart (RWR) from $v$.
Mathematically, $\procfun{v}$ is the solution to the equation
\begin{align*}
\procfun{v} = \alpha A \procfun{v} + (1-\alpha) \mathbb{1}_v,
\end{align*}
where $\alpha\in[0,1]$ is the follow-through probability 
(i.e., $1-\alpha$ is the restart probability), and 
$\mathbb{1}_v$ is the indicator function for the set~$\{v\}$.
For $\alpha \to 1$, the RWR becomes a traditional random walk; its stationary distribution is proportional to degree centrality.
For $\alpha \to 0$, the RWR remains trapped in the immediate neighborhood of $v$.
Thus, this diffusion process can be understood as interpolating between local closeness and global degree centrality.

\subsection{The \rwc Measure and its Limitations}\label{sec:no-inf-fix}
\rwc~\citep{garimella2016quantifying,garimella2018quantifying} is a widely-used structural polarization measure that informs the present work.
At its core, \rwc measures the relative exposure of a user to influential members of their own community versus those of an opposing community.
The measure assumes that influencers are high-in-degree vertices and that information spreads via random walks.

\spara{Systematic bias.}
A key critique of \rwc (as well as of other measures) is its systematic bias~\citep{salloum2022separating}.
The measure reports positive polarization scores on random \er~networks, where edges are independent, thus no true structural division exists.
Its scores on real-world networks are often preserved even after randomization with null models that preserve basic degree structure (for $d=0$ and $d=1$ using the dk-series terminology~\citep{mahadevan2006systematic}).
These results suggest \rwc is sensitive to elementary network features such as size, density, and community balance, rather than capturing only genuine structural divisions.
However, the proposed solution of standardizing the score by its values in a random graph null model~\citep{salloum2022separating} raises several crucial issues.
First, it is unclear whether vertex identities (and therefore their labels) should be fixed when computing the measure in the samples from the null model, or if the labeling (i.e., the partitioning in the broader pipeline) is assumed to be part of the measure.
Second, the solution, while widely applicable to all measures, is post-hoc. 
In addition, it requires considerable computational resources.
A more principled solution would be to understand the causes of these biases to design a measure that is unbiased by~construction.

\begin{figure}[t]
    \centering
    \begin{subfigure}{\columnwidth}
    \centering
    \includegraphics[width=\columnwidth]{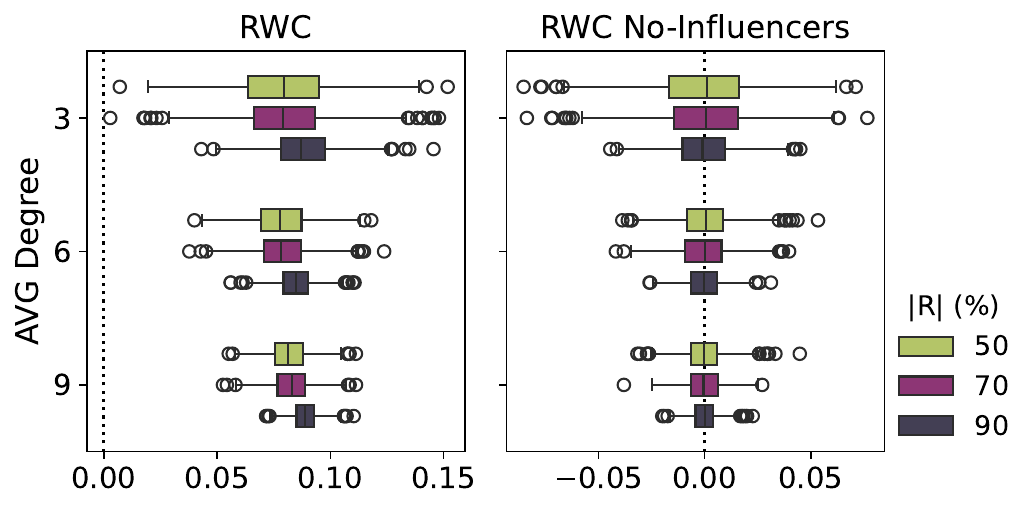}
    \end{subfigure}
    \caption{Polarization scores in \num{1000} random networks from \gnml, each with \num{10000} vertices, varying average degree and partition sizes: 50\% red--50\% blue, 70\% red--30\% blue, and 90\% red--10\% blue\@. \rwc (left) shows an unwarranted positive bias, due to overlap between the restart set and the influencers. Removing this overlap eliminates the bias, as shown for the ``no-influencers'' variant of \rwc (right).}%
    \label{fig:er-graphs-uneven}
\end{figure}

\spara{Drawbacks of \rwc.}
One of the main drawbacks of \rwc is that it yields positive values on random network null models such as \er~\citep{salloum2022separating}.
Receiving positive scores on random networks puts the measure's validity in doubt.
The issue lies in how \rwc handles its predefined set of ``influencer'' vertices.
The measure is based on the probability of a random walk starting in one community (\reds or \blues), conditioned on it having reached an influencer in a given community ($\reds^+$ or $\blues^+$).
However, a bias is introduced when the walk's restart set overlaps with the target influencer set.
This overlap artificially inflates the probability of a walk ``staying'' within its own community, as a walk starting from an influencer has already reached its destination.\footnote{In the original formulation, the set of influencers was negligible compared to the network ($10$ in networks with thousands of vertices), so the bias was not apparent. In smaller networks, or when using more influencers, this bias can be substantial.}
As shown in \Cref{fig:er-graphs-uneven}, a simple fix is to exclude influencers from the restart set, which corrects the bias for random graphs.
Although this ``no-influencers'' variant works in this specific case, it is neither a principled nor a practical solution.
This is because ($i$) there is no a priori way to define the set of influencers within each community, and ($ii$) there is no principled criterion for choosing how many influencers to include in the set.
These limitations motivate our work: to design a measure that dispenses with the arbitrary notion of influencers and treats all vertices equally~\citep{bakshy2011everyones}, leading to a more robust and theoretically sound assessment of structural polarization.

%% file: sections/algo.tex
\section{\rwcpp: Improved Polarization Measure}\label{sec:rwcpp}

This section introduces \rwcpp, our new, principled measure for structural polarization.
Similarly to previous polarization measures~\citep{garimella2016quantifying,garimella2018quantifying}, our measure 
employs as a core component a diffusion process over the network. 
In particular, we quantify how the probability mass of the diffusion process 
reaches the network communities when starting from different source vertices.
\rwcpp is designed to eliminate the biases of existing measures by removing the concept of ``influencers'' and instead considering every vertex as a potential source of diffusion.
We start by defining a general diffusion process.

\spara{The diffusion process.}
For each vertex $v \in V$, consider a diffusion process starting from $v$,
where the non-negative function $\procfun{v}: V \to \mathbb{R}^+$ (see
\cref{sec:notation}) measures how information originating from $v$ spreads.
The family of functions ${(\procfun{v})}_{v \in V}$ is a parameter of our framework.
An effective choice for $\procfun{v}$, which we adopt in the remainder of this paper, 
is the random walk with restart (RWR) vector with restart at $v$.
This choice connects our measure to the methodology of \rwc, 
where RWR is used to approximate how a user's endorsement is distributed across
the network. However, other choices of $\procfun{v}$ are possible.

For a diffusion starting from a vertex $v \in V$,
since we know that $v$ is the source of the diffusion, we want to set its score equal to zero.
Thus, to continue working with probabilities, 
for each $v \in V$, we define a probability distribution $\procprobfun{v}$ over $V$, such that
\begin{align*}
  \procprob{v}{w} \doteq \frac{\procfun{v}(w)}{\sum_{u \in V \setminus \set{v}} \procfun{v}(u)} 
  \text{~for~} w \neq v, \text{~and~} \procprob{v}{v} \doteq 0.
\end{align*}

Now, consider a thought experiment to model information flow in the network.
First, we pick a ``source'' vertex \source uniformly at random from $V$.
Then, we pick a ``target'' vertex \target by sampling from the distribution \procprobfun{\source}.
This two-step process defines a joint probability distribution over all ordered pairs of distinct vertices $(v,w) \in V \times V$.

Let us now define a function $h_Q(v)$ 
as the probability that the source $\source$ of a diffusion reaching $v$ belongs to community $Q$.
Let, for any $Q \in \set{\reds,\blues}$ and $v \in V$,
\begin{align}\label{eq:hfun}
  h_Q(v) \doteq \Pr_{\source,\target}(\source \in Q \mid \target = v).
\end{align}
We refer to $h_Q(v)$ as the \emph{exposure} of a single vertex $v$ 
to a community~$Q \in \set{\reds, \blues}$.
Using Bayes' theorem, we can express $h_Q(v)$ as a function of 
the probability distributions $\procprobfun{v}$:
\begin{align*}
    \Pr_{\source,\target}(\source \in Q \mid \target = v) 
      &= \frac{\Pr(\target = v \mid \source \in Q) \Pr(\source \in Q)}{\Pr(\target = v)} \\
    &= \frac{ \left(\frac{1}{\sizeof{Q}} \sum_{w \in Q} \procprob{w}{v}\right) \frac{\sizeof{Q}}{n} }{ \frac{1}{n} \sum_{w \in V} \procprob{w}{v} } 
    = \frac{\sum_{w \in Q} \procprob{w}{v}}{\sum_{w \in V} \procprob{w}{v}}.
\end{align*}
Function $h_{\community}$ depends directly on the user-specified family~${(\procfun{v})}_{v \in V}$.

\spara{Example.}
In a political context, $h_{\reds}(v)$ represents the probability that a piece of content or information reaching user $v$ originated from the community \reds.
A comparatively higher value of $h_{\reds}(v)$ suggests that $v$ is overly exposed to ``red opinions'' as they preferentially endorse that community.
Conversely, if $h_{\reds}(v)$ and $h_{\blues}(v)$ are balanced, user $v$ has a more diverse information diet.

\smallskip
Finally, we define a scoring function $\ell(Q, v)$ 
capturing whether the exposure of vertex $v$ to community $Q$ is a sign of polarization.
\begin{align}\label{eq:ellfun}
  \ell(Q,v) \doteq \begin{cases}
    h_{Q}(v) & \text{if } v \in Q, \\
    -h_{Q}(v) & \text{if } v \not\in Q.
  \end{cases}
\end{align}
The intuition is that, when considering the exposure of a source of the same color as the target $v$ 
(i.e., $v\in Q$), a high probability $h_{Q}(v)$ is a sign of structural homophily and contributes positively to polarization.
When instead the source is of the opposite color ($v\not\in Q$)
a high probability $h_Q(v)$ is a sign of cross-cutting exposure and contributes negatively.

\spara{A probing process for polarization.}
While the $h_Q(v)$ functions describe exposure at the vertex level, 
a network-level polarization measure requires us to aggregate these values in a principled way.
To do this, we define a \emph{probing process} that samples three random variables: 
a target community $\community_\target$, 
a probe target vertex $\probe$, and 
a source community $\community_\source$.
Formally,
\begin{squishlist}
    \item $\community_\target$ is a color chosen uniformly at random from $\set{\reds, \blues}$.
    \item $\probe$ is a vertex chosen uniformly at random from community $\community_\target$.
    \item $\community_\source$ is a community chosen from $\set{\reds, \blues}$ with a probability conditional on \probe and defined as 
\end{squishlist}
\[
  \Pr(\community_\source = \reds \mid \probe)  \doteq \frac{\sizeof{\blues \setminus \set{\probe}}}{n-1}, \text{ and }
  \Pr(\community_\source = \blues \mid \probe) \doteq \frac{\sizeof{\reds \setminus \set{\probe}}}{n-1}.
\]

\epara{Intuition.}
As we mentioned before, 
this process is designed to test for polarization systematically without needing to define influencers.
First, we select a target vertex \probe without bias toward the larger community 
(by picking a color $\community_\target$ uniformly in $\{R,B\}$).
Then, when we consider the origin of information, we choose a source community $\community_\source$ 
with probability proportional to the size of the \emph{other} community.
While this may seem counterintuitive, it serves 
to give more weight to cross-community influence in unbalanced communities, 
where otherwise, the smaller group could be drowned out.
This design is justified by the desirable properties of the resulting measure, such as producing a zero score for a fully connected clique regardless of the community sizes, as shown in \Cref{sec:analysis}.

\spara{The proposed polarization measure.}
Our new polarization measure 
combines the scoring function $\ell(Q, v)$ and the probing process defined above.
In particular, \rwcpp is defined as the expected value of the scoring function $\ell(Q, v)$ over the probing~process:
\begin{equation}\label{eq:rwcpp}
  \rwcpp = \mathbb{E}_{\community_\target,\probe,\community_\source}\left[\ell(\community_\source,\probe)\right].
\end{equation}

\epara{Intuition for \rwcpp.}
In essence, \rwcpp is the average score of how likely a vertex is to receive information from a given community. 
This average is over the distribution defined by our probing process.


\spara{Properties and expanded formulation.}
We now expand the expression for \rwcpp to analyze its properties.
By the law of total expectation, we can unroll the expectation in \Cref{eq:rwcpp}.
\begin{align*}
  \rwcpp = \frac{1}{2}\sum_{\community \in \set{\reds,\blues}}
  \mathbb{E}_{\probe,\community_\source}\left[\ell(\community_\source,\probe) \mid \community_\target=\community\right].
\end{align*}

Applying the law of total expectation again, we get
\begin{align*}
  \rwcpp &= \frac{1}{2\sizeof{\reds}} \sum_{v \in \reds} \mathbb{E}_{\community_\source}\left[\ell(\community_\source,v) \mid \probe=v, \community_\target=\reds\right]\\
  &+ \frac{1}{2 \sizeof{\blues}} \sum_{v \in \blues} \mathbb{E}_{\community_\source}\left[\ell(\community_\source,v) \mid \probe=v, \community_\target=\blues\right].
\end{align*}

Using the definition of conditional expectation, and the probability distribution for $\community_\source$, 
we obtain
\begin{align*}
  \rwcpp =& \frac{1}{2\sizeof{\reds}} \sum_{v \in \reds}
  \left(\frac{\sizeof{\blues}}{n-1} h_{\reds}(v) -
  \frac{\sizeof{\reds}-1}{n-1} h_{\blues}(v) \right)\\
  +& \frac{1}{2\sizeof{\blues}} \sum_{v \in
  \blues}\left(\frac{\sizeof{\reds}}{n-1} h_{\blues}(v) -
  \frac{\sizeof{\blues}-1}{n-1} h_{\reds}(v) \right).
\end{align*}

Combining \Cref{eq:hfun} with the previous equation, and minimally rearranging the terms, 
we get
\begin{align}
  \label{eq:rwcppderiv}
  \rwcpp &= \\
  & \frac{1}{2\sizeof{\reds}} \left( \frac{\sizeof{\blues}}{n-1} \sum_{v
  \in \reds}
  \frac{\sum_{w\in \reds}  \procprob{w}{v}}{\sum_{w \in V} \procprob{w}{v}} 
  \notag
  - 
  \frac{\sizeof{\reds}-1}{n-1} \sum_{v \in \reds}
  \frac{\sum_{w \in \blues} \procprob{w}{v}}{\sum_{w \in V} \procprob{w}{v}} \right)\\
  \notag
  +& \frac{1}{2\sizeof{\blues}} \left( \frac{\sizeof{\reds}}{n-1} \sum_{v \in \blues}
  \frac{\sum_{w \in \blues} \procprob{w}{v}}{\sum_{w \in V} \procprob{w}{v}} 
  \notag
  - 
  \frac{\sizeof{\blues}-1}{n-1} \sum_{v \in \blues}
  \frac{\sum_{w \in \reds} \procprob{w}{v}}{\sum_{w \in V} \procprob{w}{v}} \right).
\end{align}
Beyond its analytical utility for studying the properties of the \rwcpp measure, 
\Cref{eq:rwcppderiv} also provides a direct path for computation.
The equation shows that calculating \rwcpp boils down to computing four aggregate diffusion scores: within-community ($\reds \to \reds$, $\blues \to \blues$) and cross-community ($\blues \to \reds$, $\reds \to \blues$).
When using random walk with restart (RWR) for the diffusion scores $\procprob{v}{w}$, 
it is required to compute the RWR vector restarting from each vertex~$w \in V$.
The values of these vectors at each target vertex $v$ are then summed according to the community memberships of $v$ and $w$.
While computationally intensive, it is possible to use sampling-based approximations to speed up the computation.

\smallskip
\epara{Range.}
The range of \rwcpp is $(-\frac{1}{2} \frac{n-2}{n-1}, \frac{1}{2} \frac{n}{n-1})$.
The maximum is attained when diffusions from a community only reach vertices within that same community, the hallmark of extreme structural polarization.
The minimum is attained when diffusions only reach vertices of the opposite community.
A key advantage of \rwcpp is that negative values are clearly interpretable: they indicate that, on average, the network structure promotes cross-community diffusion more than within-community diffusion.
This is a more intuitive interpretation than for previously proposed measures.

To see why, assume that for every $v \in \reds$ it is
$
  \sum_{w \in \reds} \procprob{w}{v} = 1,
$
and for every $v \in \blues$ it is
$
  \sum_{w \in \blues} \procprob{w}{v} = 1.
$
It is easy to see that these assumptions maximize the value of \rwcpp to be
\[
  \frac{1}{2} \frac{n}{n-1} \to \frac{1}{2}, ~\text{ as }~ n \to +\infty.
\]

The minimum value for \rwcpp is
\[
  -\frac{1}{2} \frac{n-2}{n-1} \to -\frac{1}{2}, ~\text{ as }~ n \to +\infty,
\]
attained when, for every $v \in \reds$ it is
$
  \sum_{w \in \blues} \procprob{w}{v} = 1,
$
and for every $v \in \blues$ it is
$
  \sum_{w \in \reds} \procprob{w}{v} = 1.
$

\smallskip
\epara{More than two colors.}
It is possible to extend the \rwcpp definition to the general case of $k \ge 2$ colors.
Reusing the same notation as above, we need to define the following random variables:
\begin{squishlist}
  \item $\community_\target$, a color chosen u.a.r. from the set of
    possible~colors;
  \item $\probe$, a vertex chosen u.a.r. from the community~$\community_\target$;
  \item $\community_\source$, a set of vertices chosen from the bi-partition
    $\set{\community_\target, V\setminus \community_\target}$,
    according to the following conditional probabilities
    \begin{align*}
      \Pr(\community_\source = \community_\target \mid \community_\target) 
        &= \frac{\sizeof{V \setminus \community_\target}}{n - 1}, \text{ and}\\
      \Pr(\community_\source = V \setminus \community_\target\mid \community_\target) &=
      \frac{\sizeof{\community_\target} - 1}{n - 1}.
    \end{align*}


\end{squishlist}
Now, for any $v$ and possible community (i.e., color) $Q$, define
\begin{align*}
  \ell(Q, v) \doteq
  \begin{cases}
    h_{\col{v}}(v) & \text{if } Q = \col{v},\\
    -\left(1- h_{\col{v}}(v)\right) & \text{otherwise}.\\
  \end{cases}
\end{align*}

Finally, we define
\begin{align}
  \rwcpp \doteq
  \mathbb{E}_{\community_\target,\probe,\community_\source}\left[\ell(\community_\source,\probe)\right].
\end{align}
With $k \ge 2$ colors, the range of the \rwcpp is roughly $(-\sfrac{1}{k}\,, \sfrac{1}{k})$.
These definitions collapse to the previous ones when $k=2$.

%% file: sections/analysis.tex
\section{Analysis on Synthetic Data}\label{sec:analysis}

In this section, we prove that $\rwcpp$ behaves as desired on a set of reference
network topologies and network null models.
As presented in the previous section, the \rwcpp measure assumes a diffusion process over the network.
In this section, for concreteness, we instantiate the generic \procfun{v} functions using a random walk with restart (RWR) with restart probability $1 - \alpha$ as diffusion process.

In the following charts, to compare the various polarization measures while respecting their semantics, we normalized their values to a common interval.
Measures that capture both the magnitude and direction of polarization are rescaled to the interval $[-1, 1]$, preserving the neutral point at $0$, whereas measures that quantify only the magnitude of polarization are normalized to $[0, 1]$ to avoid introducing artificial directionality.
Finally, for the experiments in this section, we set the number of influencers in \rwc to $10$, and in its adaptive variant (ARWC) to $10\%$ of the network’s vertices.

\subsection{Reference Network Topologies}

We consider three types of networks: \emph{cliques}, \emph{color-alternating cycles},
and \emph{monochromatic-splittable networks} (defined below). 
These specific networks have a simple enough structure that enables deriving the exact solution of the $\rwcpp$ polarization measure.



\begin{figure}[t]
    \centering
    \includegraphics[width=.8\linewidth]{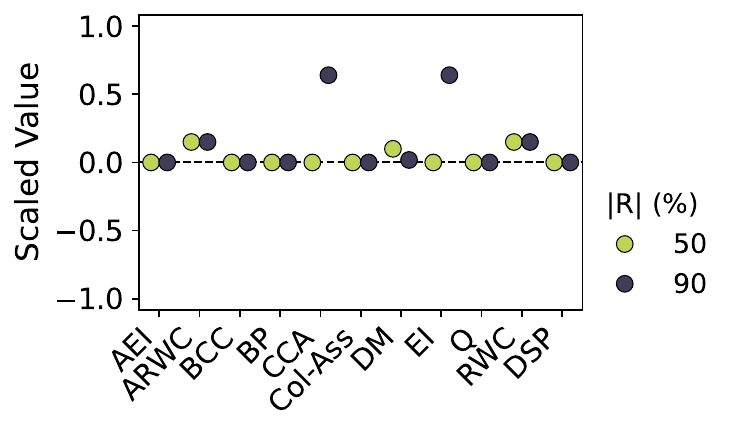}
    \caption{Polarization in a bi-colored clique with $5000$ vertices and partitions of different sizes: 50\% red--50\% blue, and 90\% red--10\% blue. The dashed line denotes the desired value of $0$.}%
      \label{fig:clique_exp}
\end{figure}

\spara{Clique.}
A clique of $n$ vertices, arbitrarily partitioned into $\reds$ and $\blues$,
possibly with $\sizeof{\reds} \neq \sizeof{\blues}$, should exhibit no structural
polarization: every vertex is connected to every other vertex, so every vertex has
the same opportunities of accessing content (or interacting with vertices) from its
own partition as it does from the other partition.

It holds that $\procprob{v}{w}=1/(n-1)$ for every $v \in  V$ and every $w \in V
\setminus \set{v}$, 
and $\procprob{v}{v}=0$ as required (see \Cref{sec:rwcpp}). The denominators of
\Cref{eq:rwcppderiv} are therefore all equal to $1$, while the four numerators
are, in order, $(\sizeof{\reds}-1)/(n-1)$, $\sizeof{\blues}/(n-1)$,
$(\sizeof{\blues}-1)/(n-1)$, and $\sizeof{\reds}/(n-1)$.
Thus, we can express $\rwcpp$ as
\begin{align*}
  \rwcpp =& \frac{1}{2\sizeof{\reds}} \left( \frac{\sizeof{\blues}}{n-1}
  \sizeof{\reds} \frac{\sizeof{\reds}-1}{n-1} - \frac{\sizeof{\reds}-1}{n-1}
  \sizeof{\reds} \frac{\sizeof{\blues}}{n-1} \right)\\
  &+ \frac{1}{2\sizeof{\blues}} \left( \frac{\sizeof{\reds}}{n-1}
  \sizeof{\blues} \frac{\sizeof{\blues}-1}{n-1} -
  \frac{\sizeof{\blues}-1}{n-1} \sizeof{\blues} \frac{\sizeof{\reds}}{n-1} \right) = 0 \enspace.
\end{align*}

\Cref{fig:clique_exp} shows the polarization values in a bi-colored clique with $5000$ vertices and two partition ratios: one balanced (50\% red, 50\% blue) and one unbalanced (90\% red, 10\% blue). We observe that \rwc and its adaptive variant \arwc exhibit a positive bias. 
In contrast, the other measures yield the desired polarization score of $0$, except for DM, which also presents a positive bias on balanced partitions,
and EI and CCA, which are sensitive to community size imbalance and yield positive scores when the partition is unbalanced.

\spara{Color-alternating Cycle.}
A color-alternating cycle is a cycle network with (even) $n$ vertices, which are
evenly split between $\blues$ and $\reds$, and placed in an alternating fashion, 
i.e., each vertex $v$ has exactly two neighbors, both of which have a color different from its own.
A negative polarization is expected in this network, because each vertex is directly connected to only vertices of the other color, and thus it is more likely to access content (and interact with vertices) from the other color than its own.

The high level of symmetry in such a network implies that there is a positive $z =
z(n, \alpha) \in (0,1)$ such that, for every vertex $v \in V$, it holds that
$h_{\col{v}}(v) = z$ and $h_{\widebar{\col{v}}}(v) = 1 - z$.
By plugging $z$ in \Cref{eq:rwcppderiv}, we obtain
\begin{align*}
  \rwcpp =& \frac{1}{2} \left( \frac{\sizeof{\blues}}{n-1} z - \frac{\sizeof{\reds}-1}{n-1} (1-z) \right) \\
  &+ \frac{1}{2} \left(\frac{\sizeof{\reds}}{n-1} z - \frac{\sizeof{\blues}-1}{n-1} (1-z) \right) = z - \frac{1}{2}\frac{n-2}{n-1} \enspace.
\end{align*}

A lower value of $\alpha$ for \procfun{v} (a RWR that restarts more often) leads to a lower value of $z$.
As $z$ is a probability, it is bounded below by zero, thus allowing us to recover the lower
bound on the range of \rwcpp as $-{1 \over 2} {n-2 \over n-1}$ as $\alpha$ goes to zero.

\begin{figure}[t]
    \centering
    \includegraphics[width=.8\linewidth]{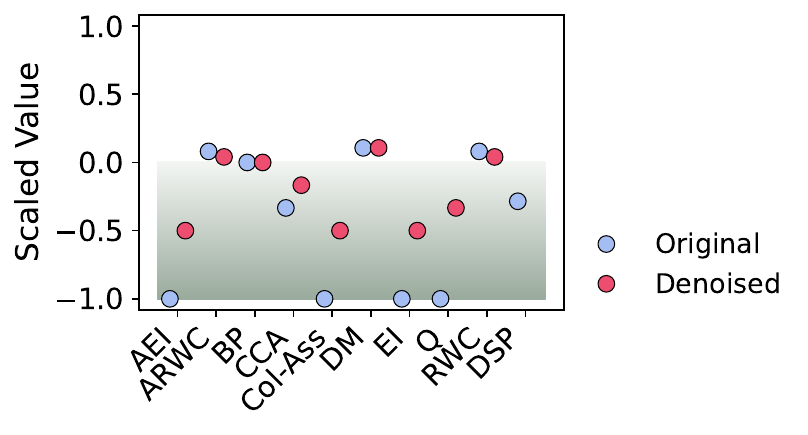}
    \caption{Polarization in a bi-colored alternating cycle with $5000$ vertices, 50\% red--50\% blue. We show the rescaled values and the rescaled denoised values computed using the $1k$-series~\citep{salloum2022separating}. The gradient area indicates the desired scores.}%
    \label{fig:alt-cyc_exp}
\end{figure}

\Cref{fig:alt-cyc_exp} shows the polarization values in a bi-colored alternating cycle with $5000$ vertices, evenly split between red and blue.
We also include the denoised~\citep{salloum2022separating} polarization scores.
The chart displays a gradient area to indicate the desired direction of polarization. 
As in the clique experiment, \rwc and \arwc display a positive bias.
We observe that the denoising often worsens the results: measures that initially yield negative values become less negative---or even turn positive---after applying the proposed correction.

\spara{Monochromatic-splittable Networks.}
A monochromatic-split\-table network is a network with a cut that splits the network into
exactly two connected components, each monochromatic, i.e., containing only
vertices of one of the two colors. As long as at least one of~$\sizeof{\reds}$
or~$\sizeof{\blues}$ is much larger than the size of the cut, these networks exhibit strongly positive polarization, because all but a few vertices of each color are connected only to vertices of the same color, and thus most interactions are between homochromatic vertices.
As $\alpha$ decreases, then the components of \Cref{eq:rwcppderiv}
\begin{align*}
  \sum_{v \in \reds} \frac{\sum_{w \in \reds} \procprob{w}{v}}{\sum_{w \in V}
  \procprob{w}{v}} \rightarrow \sizeof{R}\,,
\end{align*}
and similarly for $\blues$. The rationale is the same for the alternating cycle:
smaller $\alpha$ values lead to more importance for the immediate neighbors of
the starting vertex, which, in this case, are vertices with the same color as the
starting vertex. Thus, the value of \rwcpp tends to its maximum attainable value
$\frac{n}{2(n-1)}$, as these are exactly the conditions that we assumed when analyzing
the range of \rwcpp (recall that $\rwcpp \in \left(\sfrac{-1}{2}\,,\sfrac{1}{2}\right)$ \text{approximately}).

\begin{figure}[t]
    \centering
    \includegraphics[width=\linewidth]{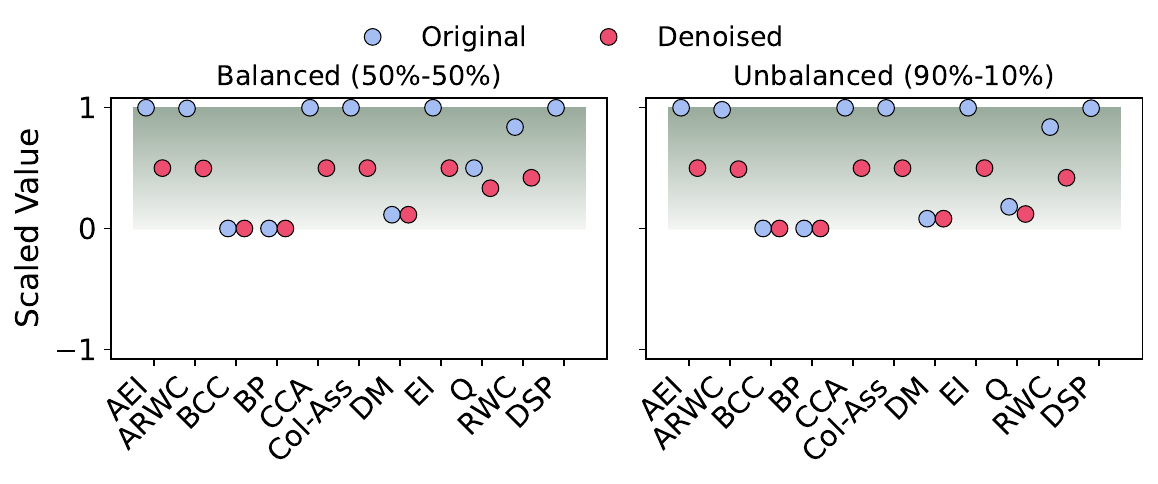}
    \caption{Polarization in a bi-colored half-split cycle with $5000$ vertices and two partition sizes: 50\% blue--50\% red (left), and 90\% red--10\% blue (right). We show the rescaled and rescaled-denoised values computed using the $1k$-series~\citep{salloum2022separating}.
    The gradient area indicates the desired scores.}%
    \label{fig:half-split-cycle-exp}
\end{figure}

\begin{figure}[t]
    \centering
    \includegraphics[width=\linewidth]{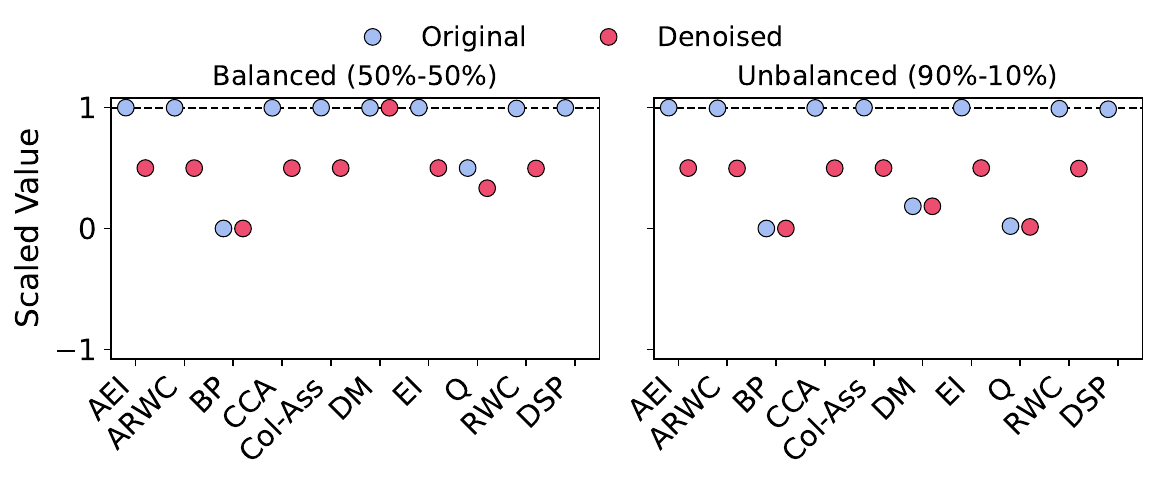}
    \caption{Polarization in a bi-colored half-split barbell network with $2000$ vertices and two partition sizes: 50\% blue--50\% red (left), and 90\% red--10\% blue (right). We show the rescaled and rescaled-denoised values computed using the $1k$-series~\citep{salloum2022separating}. The dashed line denotes the desired value of $1$.}%
    \label{fig:half-split-bar-exp}
\end{figure}

When the two connected components are of the same size (i.e., $\sizeof{\reds} =
\sizeof{\blues} = \sfrac{n}{2}$), and isomorphic, then a more precise analysis
is possible. Examples of such networks are the half-split cycle network obtained by
connecting both ends of two monochromatic chains (one red and one blue) of the same size, or
a ``barbell network'' where two monochromatic cliques of the same size are
connected by a chain network that is half of one color and half of the other color.
In these cases, symmetry implies that there are non-negative values $z_1 =
z_1(n, \alpha)$ and $z_2 = z_2(n, \alpha)$ such that
\begin{align*}
  \!\!
  \rwcpp &= \frac{1}{2\sizeof{\reds}} \left( \frac{\sizeof{\blues}}{n-1} z_1 -
  \frac{\sizeof{\reds}-1}{n-1} z_2 \right)\nonumber
  + \frac{1}{2\sizeof{\blues}} \left( \frac{\sizeof{\reds}}{n-1} z_1 - 
  \frac{\sizeof{\blues}-1}{n-1} z_2 \right)
  \nonumber\\
  =& \frac{1}{n} \left( \frac{\sizeof{\blues} + \sizeof{\reds}}{n-1} z_1 -
  \frac{\sizeof{\reds} + \sizeof{\blues} - 1 - 1}{n-1} z_2 \right)
  = \frac{1}{n-1}z_1 - \frac{1}{n} \frac{n-2}{n-1} z_2
  \enspace.
\end{align*}
It holds
\begin{align*}
  z_1 + z_2 &= \sum_{y \in \reds} \frac{\sum_{i \in \reds}
  \procprob{i}{y}}{\sum_{i \in V} \procprob{i}{y}} + \sum_{y \in \reds}
  \frac{\sum_{i \in \blues}
  \procprob{i}{y}}{\sum_{i \in V} \procprob{i}{y}} \\
  &= \sum_{y \in \reds} \frac{\sum_{i \in \reds}
  \procprob{i}{y}}{\sum_{i \in V} \procprob{i}{y}} = \sizeof{\reds} =
  \frac{n}{2} \enspace .
\end{align*}
Thus, we can express $z_2$ as $\sfrac{n}{2} - z_1$. Continuing from
\Cref{eq:rwcppderiv},
\begin{align*}
  \rwcpp &= \frac{1}{n-1}z_1 - \frac{1}{n} \frac{n-2}{n-1} \left( \frac{n}{2} - z_1 \right)
  = \frac{2}{n} z_1 - \frac{1}{2} \frac{n-2}{n-1} \enspace.
\end{align*}

The value $z_1$, which is a function of $n$ and $\alpha$, increases as $\alpha$
decrease. As $z_1$ is upper bounded by $\frac{n}{2}$, we 
recover the upper bound on the range of $\rwcpp$ as 
$\frac{1}{2}\frac{n}{n-1}$.

\Cref{fig:half-split-cycle-exp} presents the polarization values in a bi-colored half-split cycle with $5000$ vertices, for two partitioning schemes: one with equal-sized red and blue groups, and another with a 90\%-10\% split. As before, the figure includes the rescaled denoised polarization scores computed using the $1k$-series correction method. 
The gradient area highlights the desired direction of polarization. 
We observe that all measures yield positive scores for both partitioning scenarios; however, in several cases, the denoising approach reduces these values, hence worsening their performance.
\rwcpp consistently achieves the highest possible score, along with a few other measures.


\Cref{fig:half-split-bar-exp} reports results for a bi-colored half-split barbell network with $2000$ vertices, again using both balanced and 90\%--10\% partitions. 
Rescaled denoised scores are also shown. 
The black dashed line represents the desired polarization value of 1. 
Most measures correctly reach this value in the balanced case, but performance tends to degrade---particularly after denoising---in the unbalanced setting.
Overall, denoising~\citep{salloum2022separating} often moves the polarization scores away from their ideal values rather than correcting them. Across all scenarios, only two measures consistently align with the expected behavior: \rwcpp and AEI.%

%
%
%

\begin{figure}[t]
    \centering
    \includegraphics[width=.9\linewidth]{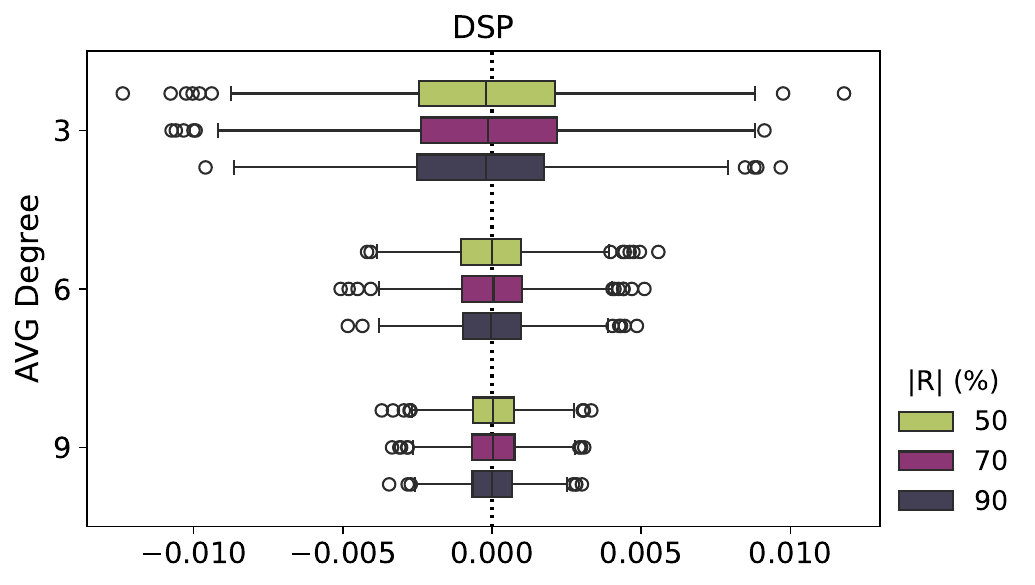}
	\caption{\rwcpp in $1000$ random networks extracted from \gnml for different average degrees $d$ and partition sizes $\ell(\reds) \in \{50\%,70\%,90\%\}$. We set $n=\num{10000}$ and $p = \sfrac{d}{n - 1}$.}%
    \label{fig:gnml_rwcpp}
\end{figure}

\begin{table}[t]
\caption{Polarization in $1000$ random networks from \gnml for different average degrees $d$ and partition sizes $\ell(\reds) \in \{50\%,70\%,90\%\}$. We set $n=\num{10000}$ and $p = \sfrac{d}{n - 1}$.}%
\label{tab:gnml}
\centering
    \resizebox{\linewidth}{!}{
    \begin{tabular}{ccrrrrrrrrrr}
    \toprule
    & & \multicolumn{10}{c}{\textbf{Metric}} \\
    \cmidrule{3-12}
    $d$ & $|R| (\%)$ & AEI & ARWC & BCC & BP & CCA & Col-Ass & DM & EI & Q & \rwc \\
    \midrule
    \multirow[c]{3}{*}{3} 
    & 50 & 0 & 0.115 & 0.298 & -0.187 & 0 & -0.001 & 0.444 & 0 & 0 & 0.079 \\
    & 70 & 0 & 0.115 & 0.299 & -0.039 & 0.161 & -0.001 & 0.194 & 0.160 & 0 & 0.080 \\
    & 90 & -0.001 & 0.115 & 0.303 & 0.260 & 0.634 & 0 & 0.050 & 0.640 & 0 & 0.088 \\
    \cmidrule{2-12}
    \multirow[c]{3}{*}{6} 
    & 50 & 0 & 0.106 & 0.115 & -0.416 & 0 & 0 & 0.264 & 0 & 0 & 0.078 \\
    & 70 & 0 & 0.106 & 0.115 & -0.184 & 0.161 & 0 & 0.076 & 0.160 & 0 & 0.079 \\
    & 90 & 0 & 0.106 & 0.115 & 0.344 & 0.640 & 0 & 0.021 & 0.640 & 0 & 0.085 \\
    \cmidrule{2-12}
    \multirow[c]{3}{*}{9}
    & 50 & 0 & 0.108 & 0.070 & -0.572 & 0 & 0 & 0.214 & 0 & 0 & 0.082 \\
    & 70 & 0 & 0.108 & 0.070 & -0.342 & 0.160 & 0 & 0.052 & 0.160 & 0 & 0.083 \\
    & 90 & 0 & 0.108 & 0.071 & 0.337 & 0.640 & 0 & 0.018 & 0.640 & 0 & 0.089 \\
    \bottomrule
    \end{tabular}
    }
\end{table}

\subsection{The \texorpdfstring{\gnml}{G(n,m,l)} Null Model}
The \gnml model is an extension of the classical \er random network $G(n,p)$, which generates a network with $n$ vertices, each pair of which is connected with probability $p$ ($0 \leq p \leq 1$).
In this extension, we introduce labels for the vertices.
Let $\mathcal{L} \doteq [\ell_1, \ell_2, \dotsc, \ell_k]$ be the list of labels.
The null model assigns the $k$ labels uniformly at random to the vertices, subject to the constraint that exactly $\mathbf{\ell}(i)$ vertices receive label $\ell_i$.

The key property of this null model is that vertex labels are independent of the network structure, or, in other words, the edge placement process does not consider vertex labels.
Thus, any observed association between labels and structural properties is purely due to chance.
For this reason, any reasonable structural polarization measure should return, on expectation, zero or near-zero values.

\Cref{fig:gnml_rwcpp} shows the distribution of \rwcpp values measured in $1000$ random networks extracted from \gnml with  $n=\num{10000}$, different average degrees (which determines $p$), and different partition skews.
The average \rwcpp score is close to $0$ with minimal deviations for all average degrees and partition skews, proving the robustness of the measure to the partition size and network density.

\Cref{tab:gnml} reports the average scaled scores for the other polarization measures.
As expected, Q yields an average score of $0$, since it uses an \er network as its null model. Consistent with previous observations, \rwc and ARWC exhibit a positive bias, as do EI, BCC, CCA, and DM. In contrast, BP shows a negative bias. Among the other measures, only AEI consistently produces the desired near-zero values across all conditions.



\subsection{The Stochastic Block Model}\label{sec:sbm}
To further test the robustness of \rwcpp, we generate random networks using a stochastic block model (SBM) with $n$ vertices assigned uniformly at random to $2$ blocks. 
The degree distribution follows a Poisson distribution, with intra-block edge probability $p$ and inter-block edge probability $q$. 
Higher values of $p$, especially when paired with lower values of $q$, correspond to higher structural polarization.
In contrast, increasing $q$ leads to greater inter-community connectivity, hence reducing structural polarization.

\begin{figure}[t]
    \centering
    \includegraphics[width=.9\linewidth]{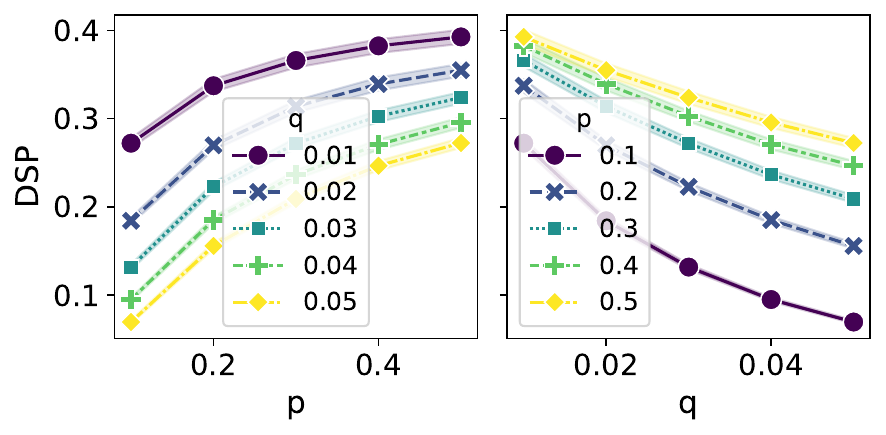}
    \caption{\rwcpp measured in $100$ networks sampled from the SBM with $1600$ vertices and $2$ blocks.}%
    \label{fig:sbm_exp}
\end{figure}

\Cref{fig:sbm_exp} shows \rwcpp for $100$ random networks with $1600$ vertices, varying $p$ and $q$. 
\rwcpp behaves as expected: it increases with denser intra-community connectivity and decreases as inter-community connectivity increases. For a fixed intra-community connectivity, lower inter-community connectivity gives higher scores.

%% file: sections/exp.tex
\section{Experiments on real-world data}\label{sec:exp}

In \cref{sec:analysis} we examined how \rwcpp behaves on our reference networks, 
where the ground-truth polarization is known.
In this section we consider real-world networks.
First, we evaluate whether \rwcpp can effectively distinguish between polarized and non-polarized networks and achieve classification performance comparable to that of existing polarization measures.
Next, we study the performance of an approximate version of \rwcpp that computes the measure considering only a sample of the network's vertices.
This approximation aims to reduce computational cost while maintaining accuracy. 
Then, we investigate the relationship between assortativity and \rwcpp, via a null model that preserves the color assortativity of the given network~\citep{preti2025polaris}.
Finally, we present a case study on political polarization using bill co-sponsorship data and roll-call voting records.
The code is available on GitHub.\footnote{\url{https://github.com/lady-bluecopper/diffusionBasedStructuralPolarization}}

\spara{Datasets.}
We consider several collections of real-world networks.
\textsc{salloum}~\citep{salloum2022separating} is a collection of $183$ polarized and non-polarized Twitter retweet networks---150 constructed from single hashtags and 33 from multiple hashtags---collected during the $2019$ Finnish Parliamentary Elections.
\textsc{congress-bill-cosp}~\citep{adler2013congress} is a set of networks based on bill co-sponsorship data in the US Congress, covering the 93rd to the 114th Congresses. 
Each edge connects two legislators, with edge weights indicating the number of times they co-sponsored a bill or joint resolution~\citep{preti2024higherorder}.
\textsc{congress-bill-roll-call}~\citep{lewis2019voteview} is a set of networks constructed from roll-call voting records in the US Senate and House for the 93rd to 114th Congresses.
As with the co-sponsorship dataset, we consider only bills and joint resolutions.
Edges connect legislators who voted identically; edge weights denote the number of such instances.

For vertex labels, for \textsc{congress-bill-cosp} and \textsc{congress-bill-roll-call}, we use the legislators' political parties (Democrat or Republican).
For \textsc{salloum}, we generate vertex labels using a graph partitioning algorithm, as commonly done in prior work~\citep{garimella2016quantifying}.
We employ the Kernighan---Lin algorithm (KLIN)~\citep{kernighan1970efficient} and METIS~\citep{karypis1997metis}.

\begin{figure}[t]
    \centering
    \includegraphics[width=.8\linewidth]{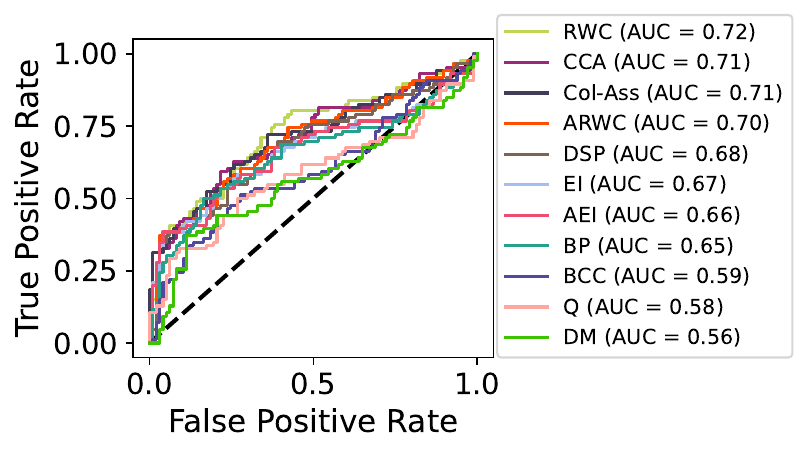}
    \caption{ROC curves and AUC values for \textsc{salloum}.}%
    \label{fig:roc_salloum}
\end{figure}

\spara{Classification performance.}
We assess how well each polarization measure distinguishes polarized from non-polarized networks using the score as the output of a probabilistic classifier~\citep{salloum2022separating}.
Each decision threshold yields a false positive and true positive rate, and varying the threshold produces a ROC curve that captures the discriminative power of each measure.
In \Cref{fig:roc_salloum} we report the unnormalized Area Under the Curve (AUC) as a summary metric of predictive accuracy.
Also in this experiment, we set the number of influencers in \rwc to $10$, and in ARWC to $10\%$ of the network’s vertices.
%
While \rwcpp is designed as a statistically principled reformulation of \rwc, the purpose of this experiment is not to demonstrate empirical superiority, but to verify that \rwcpp preserves \rwc’s practical discriminative ability.
Results show that \rwcpp achieves AUC values close to those of \rwc, suggesting that adopting a more rigorous and unbiased formulation does not compromise the predictive performance observed in earlier metrics.

\begin{figure}[tb]
    \centering
    \includegraphics[width=.47\columnwidth]{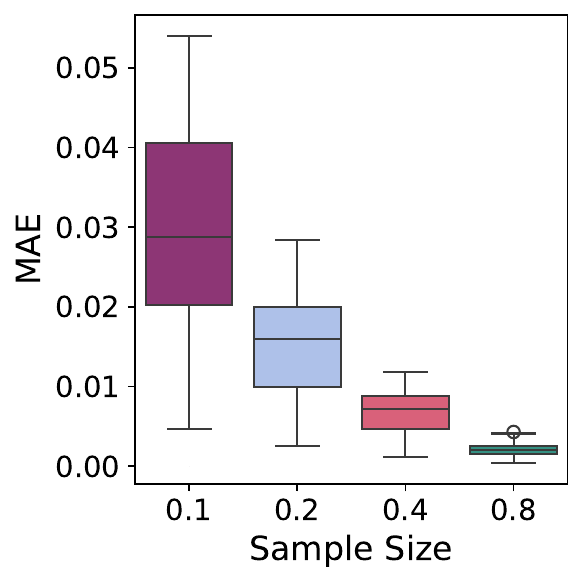}
    \hfill
    \includegraphics[width=.47\columnwidth]{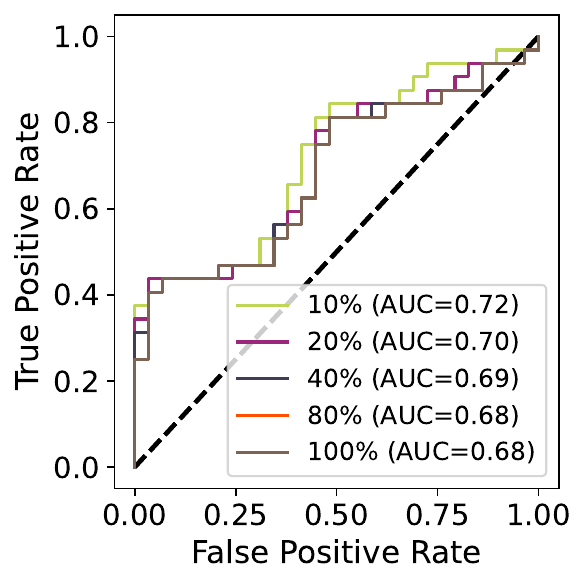}
    \caption{MAE of the approximate \rwcpp scores computed using subsets of
    vertices of different size (left); and corresponding ROC curves (right).}%
    \label{fig:approx_rwcpp}
\end{figure}

\begin{figure}[tb]
    \centering
    \includegraphics[width=\linewidth]{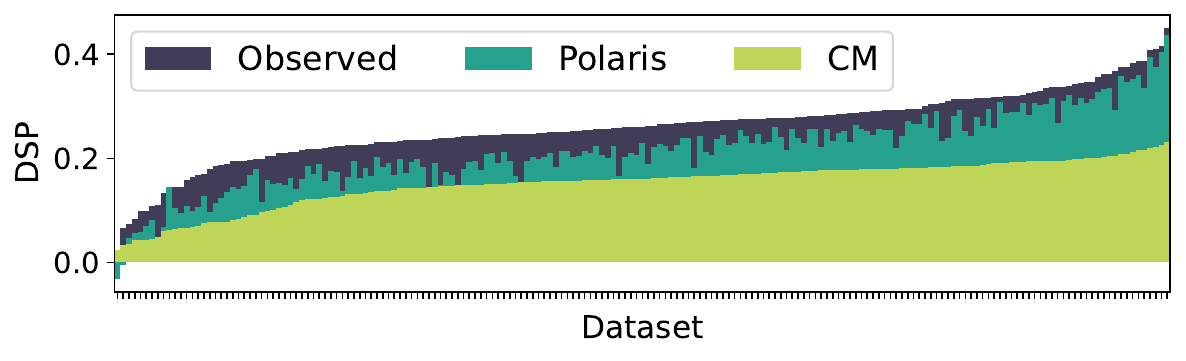}
    \caption{\rwcpp scores and average values computed over $100$ random
    networks from the configuration model (CM) and the Polaris
  model~\citep{preti2025polaris} on the \textsc{salloum} dataset.}%
    \label{fig:polaris_exp}
\end{figure}


\begin{figure*}[tb]
    \centering
    \subfloat[\textbf{\textsc{congress-bill-cosp}}]{\label{fig:congress-bill-cosp}\includegraphics[width=\columnwidth]{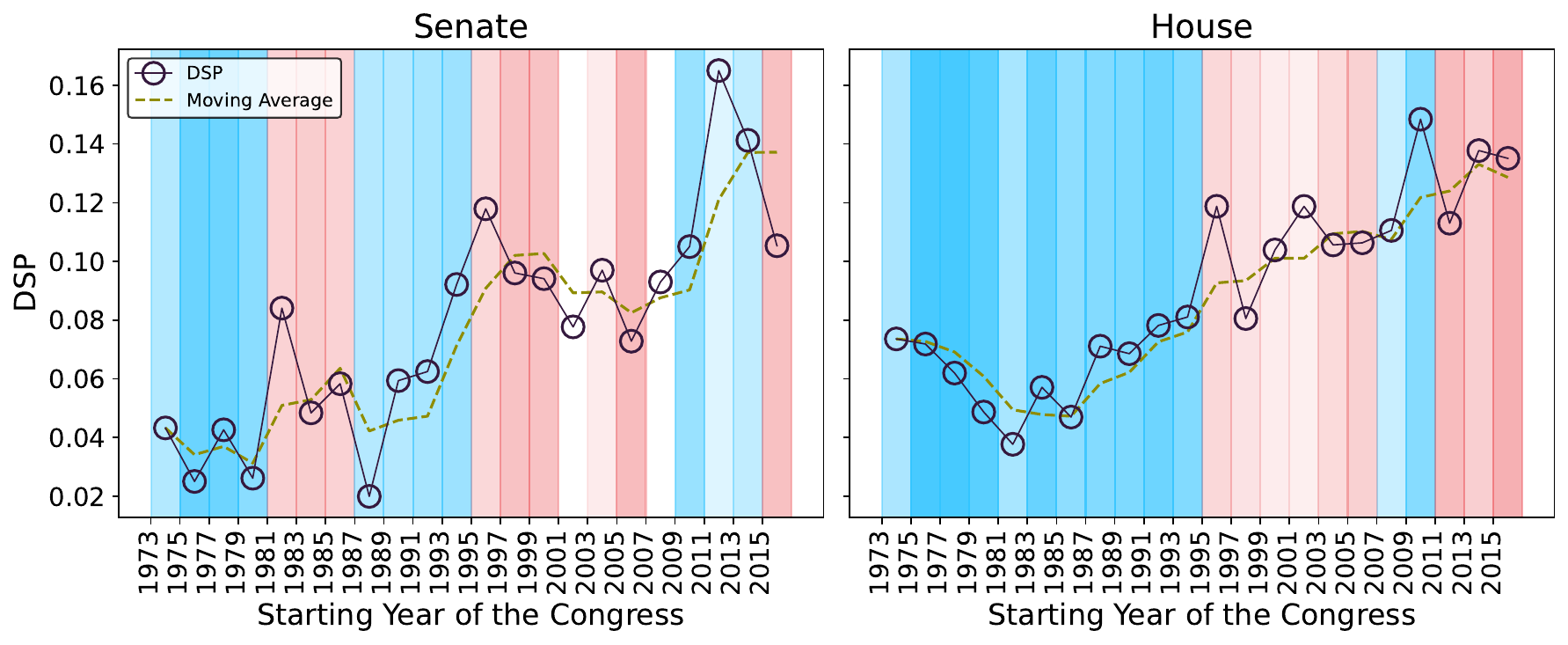}}
    \subfloat[\textbf{\textsc{congress-bill-roll-call}}]{\label{fig:congress-bill-roll-call}\includegraphics[width=\columnwidth]{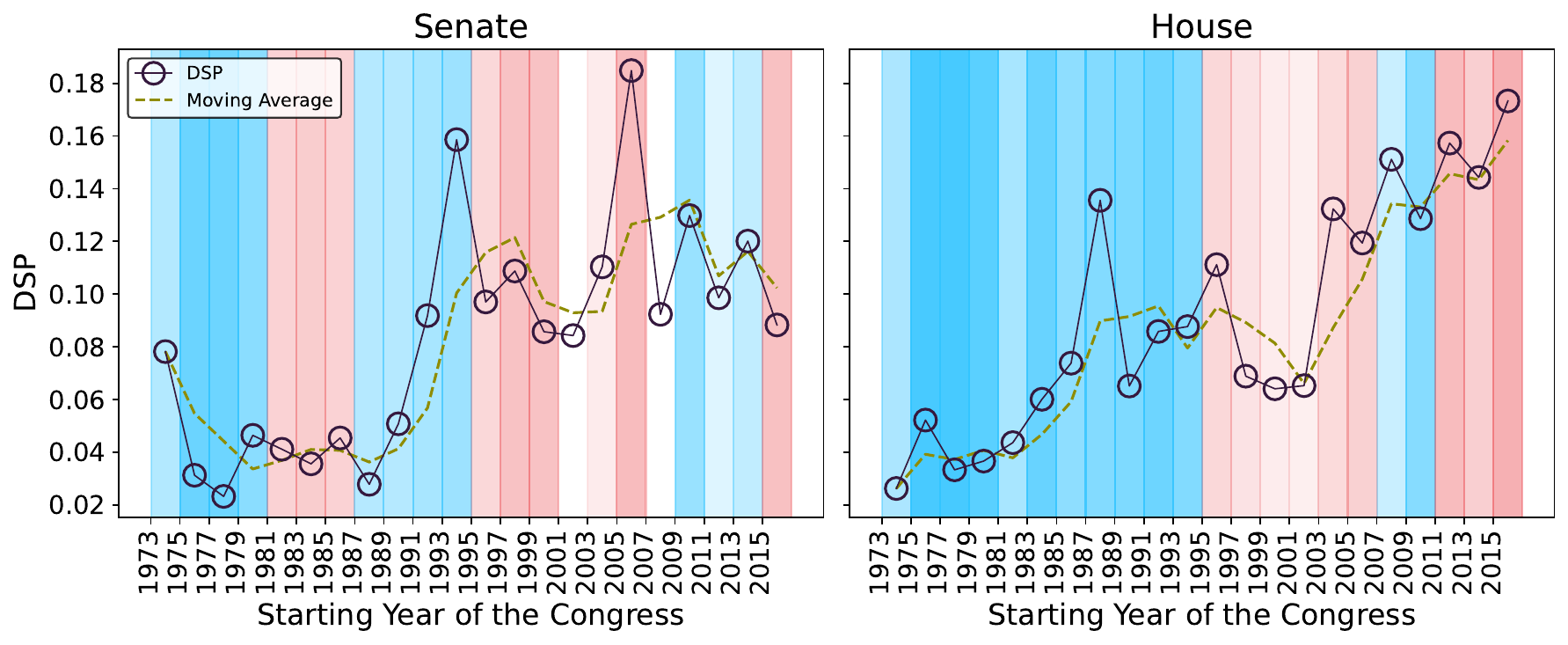}}
    \caption{\rwcpp and its 3-point moving average for both chambers across
    Congress sessions.}%
    \label{fig:congress_exp}
\end{figure*}

\spara{Approximate \rwcpp.}
To evaluate the trade-off between accuracy and efficiency in computing the \rwcpp score, we compare the exact score to a heuristic approximation obtained by calculating the summations in \Cref{eq:rwcppderiv} over a uniform random sample of the network vertices.
Computing the RWR for every vertex in the network becomes computationally expensive as the network size increases.
By considering only a subset of vertices, we aim to reduce computation time while incurring a small loss in accuracy.
In this experiment, we consider a subset of $61$ datasets from \textsc{salloum} and, for each one, extract $50$ random vertex samples with varying sample sizes.
For each sample, we compute the approximate \rwcpp score and then report the Mean Absolute Error (MAE).

In our experiments below, 
we seek to estimate the empirical number of samples required to obtain a good approximation
as a fraction of the network size.
From a theoretical point of view, 
it is plausible that the required number of samples (sample complexity) is determined by a function that grows sub\-linear\-ly with respect 
to the network size. 
We leave the question of theoretically deriving the sample complexity as future work.
\Cref{fig:approx_rwcpp} (top) shows that we need to retain at least 20\% of the vertices to achieve a good approximation of the original score.
To reduce variability in the approximation across samples, we should retain at least 40\% of the vertices.
Finally, the chart on the right shows that the ROC curves and the corresponding AUC values are roughly the same across sample sizes, indicating that even when using a small vertex set, the score has the same capability of distinguishing between polarized and non-polarized~networks.

\spara{Color assortativity vs.\ polarization.}
We compare the \rwcpp values measured on real datasets with those obtained from $100$ samples generated using Polaris~\citep{preti2025polaris}, an algorithm that samples from the ensemble of networks with the same degree sequence and color assortativity.
\Cref{fig:polaris_exp} illustrates how well \rwcpp is preserved under the Polaris null model.
The chart shows the value of \rwcpp measured in the \textsc{salloum} networks and the average values computed over $100$ random networks from two null models: the configuration model (CM) and the Polaris null model.
When the bars corresponding to the two null models are close, it indicates that color assortativity has little influence on the \rwcpp value, since both models yield similar results despite one preserving assortativity and the other not. 
On the other hand, the closer the \textsc{Polaris} bar is to the observed value, the more color assortativity can explain the polarization captured by \rwcpp.
A large gap between the observed value and the \textsc{Polaris} baseline suggests that other structural or behavioral dynamics, beyond color assortativity, are contributing to the network's polarization.

Overall, networks with higher polarization levels tend to show a stronger influence from color assortativity, as the \textsc{Polaris} baseline is closer to the observed value as polarization increases. 
Meanwhile, the influence of the degree distribution remains relatively constant across datasets, as indicated by the consistent gap between the CM baseline and the observed values.
Nonetheless, the differences between the observed and \textsc{Polaris} values in less polarized networks suggest that color assortativity alone is insufficient to fully account for the observed polarization, especially when the score is low.

\spara{Polarization in the US Congress.}
We study how political polarization evolves and whether it correlates with the political control of the chambers of the US Congress~\citep{preti2024higherorder}. 
We construct two sets of bi-colored networks: one based on bill co-sponsorship
data (\textsc{congress-bill-cosp}) and the other on roll-call voting data
(\textsc{congress-bill-roll-call}), for each chamber of Congress (Senate and
House) and each session of Congress from the 93rd to the 114th\@.
To reduce noise from widely co-sponsored legislation, we discard bills with more than 25 co-sponsors.
We focus on two legislative types, i.e., Bills and Joint Resolutions, as both require passage by both chambers and the President's signature to become law.
These types are most likely to reflect strategic behavior relevant to polarization analysis.

We compute \rwcpp on each of these networks.
In this setting, the RWR vectors used in the computation of \rwcpp incorporate edge weights, i.e., co-sponsorship/co-voting frequency influences the random walk behavior.
This choice enables the polarization measure to capture stronger ties.

\Cref{fig:congress_exp} shows the \rwcpp score for each session of Congress, separately for the Senate (top) and the House (bottom).
The background color of each bar indicates which party held the majority in each chamber during that session, where darker colors correspond to stronger majorities.
The figure also reports a moving average (window size 3) to smooth trends across Congress sessions (dashed line).
Both co-sponsorship and roll-call voting data provide valuable yet distinct insights into the legislators' preferences.
Roll-call votes are recorded decisions that offer a clear signal of preference.
In contrast, bill co-sponsorship is a voluntary activity that indicates a positive disposition toward a bill; however, the absence of co-sponsorship has no clear interpretation.
Despite these differences, previous work~\citep{aleman2009comparing} showed that co-sponsorship data can produce estimates of ideal points that are highly correlated with those derived from roll-call votes.
Consistently, we observe correlated \rwcpp scores between the two types of congress networks, especially in the House, where both Pearson (0.73) and Spearman (0.70) are high.
In the Senate, while Pearson is lower (0.59), Spearman remains high (0.74), indicating a strong agreement in trend.

Similarly, our findings on the bill-cosponsorship networks align with previous studies~\citep{hohmann2023quantifying}, as we also observe low polarization levels (\rwcpp remains below 0.16 across all sessions).
This result is expected, as \rwcpp is a structural measure, and thus, will yield high values only when the network presents a strong structural division. 
Nonetheless, consistent with previous research, we observe a general increase in polarization over time~\citep{garimella2017effect}.

%% file: sections/conclusion.tex
\section{Conclusion}\label{sec:conclusion}
We introduce \rwcpp, a statistically principled structural polarization measure,
based on a probing process of information diffusion spreading as random walks
from each vertex. It does not suffer from the biases exhibited by previous
measures, as we show by analytically deriving \rwcpp values on families of
reference networks. The results of our experimental evaluation highlight how
\rwcpp can reliably differentiate between polarized and non-polarized networks.
Interesting directions for future work include deriving additional
properties of \rwcpp on more reference classes, and using \rwcpp to
recommend edges to add to the network to reduce
polarization~\citep{garimella2017reducing,haddadan2021repbublik}.

%% file: sections/ethics.tex
\section*{Ethical Considerations}
\label{sec:ethics}

The proposed \rwcpp measure is designed as a diagnostic tool to help researchers, platforms, and policymakers better understand and identify structural polarization, a phenomenon with significant societal consequences.
However, like any technology that measures social dynamics, it can be used for purposes other than its intended goal.
A primary concern is the potential for misuse by malicious actors.
For instance, a sophisticated entity could leverage \rwcpp to identify highly polarized and vulnerable communities to more effectively target them with tailored misinformation, deepening societal divisions for political or financial gain.
Similarly, an authoritarian regime could use this measure to detect and justify the suppression of dissenting groups by labeling their cohesive, self-contained networks as a source of dangerous polarization, thereby using the metric as a pretext for censorship or control.

Beyond malicious use, harms can arise even when \rwcpp is used as intended.
A quantitative score, no matter how statistically principled, is an abstraction of complex social reality.
There is a risk that a high polarization score could be used to stigmatize a community, leading to oversimplified judgments and a lack of a nuanced, qualitative understanding of that community's context.
For example, a marginalized group may form a dense, insular network for mutual support, which could be misread as structural polarization.
Furthermore, interventions designed to decrease a network's \rwcpp score could have unintended negative consequences if applied naively.
An automated system aimed at ``depolarizing'' a network might suggest interventions that disrupt vital community ties.
To mitigate these risks, we stress that \rwcpp should be used as a diagnostic aid, not a definitive verdict.
Its findings must be interpreted within a broader socio-political context, and any interventions based on it should be human-centric and subject to ethical review to ensure they do not harm the communities they intend to help.

%% file: sections/appendix.tex



\section{Structural Polarization Measures}

\spara{The Betweenness Centrality Controversy (BCC)}~\citep{garimella2018quantifying} measure compares the edge betweenness centrality of boundary and non-boundary links, by computing the KL-divergence $d_{\text{KL}}$ between the two corresponding distributions:
\[
\mathsf{BCC} \doteq 1 - \exp^{-d_{\text{KL}}}\;.
\]
The intuition is that if the two communities are strongly separated, then the links on the boundary are expected to have high edge betweenness centralities.

\spara{The Boundary Polarization (BP)}~\citep{guerra2013measure} compares the concentration of high-degree nodes within the communities ($I$) and their concentration in the boundary ($C$):
\[
\mathsf{BP} \doteq \frac{1}{|C|}\sum_{s \in C}\frac{d_I(s)}{d_C(s) + d_I(s)} - 0.5\;,
\]
where $d_C(s)$ is the degree of $s$ restricted to neighbors in $C$.
The intuition is that the further away authoritative users are from the boundary, the larger the amount of polarization present in the network.

\spara{The Cross-community Affinity (CCA}~\citep{nair2024cross} is a heterophily-based polarization metric designed for networks with two or more ideological communities.
For each node $i$, let $s(i)$ denote its community, $C$ the set of communities, and $w(s(i),c)$ the ideological distance between communities $s(i)$ and $c$ ($-1$ when $s(i) = c$ and $1$ otherwise).  
The cross-community affinity of node $i$ is defined as:
\[
\mathrm{CCA}(i) = \mathrm{DNE}(i) + \alpha \cdot \mathrm{INE}(i),
\]
where $\mathrm{DNE}(i)$ is the \emph{direct neighbor effect},
\[
\mathrm{DNE}(i)
= \sum_{c \in C} w(s(i),c) \cdot \frac{k_c(i)}{k(i)},
\]
with $k_c(i)$ the number of neighbors of $i$ in community $c$ and
$k(i)$ the total degree of $i$.
The term $\mathrm{INE}(i)$ is the \emph{indirect neighbor effect}:
\[
\mathrm{INE}(i)
= \frac{1}{|\mathrm{CN}(i)|}
  \sum_{c \in \mathrm{CN}(i)} \mathrm{ANE}_c(i),
\]
where $\mathrm{CN}(i)$ is the set of communities appearing in $i$'s
neighborhood and $\mathrm{ANE}_c(i)$ is the average neighbor effect
exerted by $i$’s neighbors in community $c$:
\[
\mathrm{ANE}_c(i)
= \frac{1}{k_c(i)}
  \sum_{j \in N_c(i)} \mathrm{NE}(j,i),
\]
with $N_c(i)$ the neighbors of $i$ in community $c$.  The effect of a
specific neighbor $j$ on $i$ is:
\[
\mathrm{NE}(j,i)
= \sum_{g \in C}
    w(s(i),g)\cdot
    \frac{k_g(j) - \delta_{g,s(i)}}{k(j)-1},
\]
where $\delta_{g,s(i)}$ is the Kronecker delta used to exclude $i$
from $j$'s neighborhood.  The parameter $\alpha = 1/h$ discounts
indirect effects with hop distance $h$ ($h=2$ in our experiments).

The Cross-community affinity of the network is the negative average node cross-community affinity:
\[
\mathrm{CCA} = -\frac{1}{|N|}\sum_{i \in N} \mathrm{CCA}(i).
\]

\spara{The Color Assortativity (Col-Ass)}~\citep{newman2003mixing} measures the tendency of nodes to connect to others of the same color.
Let $e_{cc'}$ be the fraction of edges from nodes of colors $c$ to nodes of color $c'$, $a_c = \sum_{c'}e_{cc'}$ be the fraction of edges with sources of color $c$, and $b_c = \sum_{c'}e_{c'c}$ be the fraction of edges with destinations of color $c$ (when the graph is undirected, $a_c = b_c$). Color assortativity is defined using Newman’s assortativity coefficient for categorical attributes as follows:
\[
\mathsf{Col-Ass} \doteq \frac{\sum_c e_{cc} - \sum_c a_c b_c}{1 - \sum_c a_c b_c}\;.
\]

A value of $1$ indicates perfect same-color mixing, $0$ indicates no assortative preference, and negative values indicate disassortative mixing.

\spara{The Dipole Moment (DP)}~\citep{morales2015measuring} measure applies label propagation for quantifying the distance between the top-k influencers of each community, which are assigned the extreme opinion scores of -1 or 1.
Let $gc^+$ and $gc^-$ denote the average of positive and negative opinion scores. 
The DP measure is a function of the distance between the means of the opposite opinion score distributions, rescaled to penalize differences in the community sizes: 
\[
\mathsf{DM} \doteq \left(1 - \frac{n^+ - n^-}{n^+ + n^-}\right) \frac{|gc^+ - gc^-|}{2}\;.
\]

\spara{The Krackhardt E/I Ratio (EI)}~\citep{krackhardt1988informal} is defined as the relative density of intra-edges compared to the number of inter-edges:
\[
\mathsf{EI} \doteq \frac{\text{EL} - \text{IL}}{\text{EL} + \text{IL}}\;,
\]
where EL is the set of edges with endpoints belonging to different communities, and IL is the set of edges with endpoints belonging to the same community.

\spara{The Adaptive E/I Index (AEI)}~\citep{chen2021polarization} extends the EI measure to account for communities with different sizes:
\[
\mathsf{AEI} \doteq \frac{\sigma_{aa} + \sigma_{bb} - 2*\sigma_{ab}}{\sigma_{aa} + \sigma_{bb} + 2*\sigma_{ab}}\;,
\]
where $\sigma_{aa}$ is the ratio between the number of intra-edges in community $a$ and the number of potential intra-edges, and $\sigma_{ab}$ is the ratio of the number of inter-edges between community $a$ and $b$ and the number of potential inter-edges.

\spara{Modularity (Q)}~\citep{waugh2009party} compares the connectivity of the communities to that observable in random graphs extracted from the configuration model:
\[
\mathsf{Q} \doteq \frac{1}{2|E|}\sum_{i,j \in V}\left(A_{i,j} - \frac{d(i) d(j)}{2|E|}\right)\delta(i,j)\;,
\]
where $A$ is the adjacency matrix of the graph and $\delta(i,j)$ equals one only when $i$ and $j$ belong to the same community.

\subsection{Scope of Included and Excluded Measures}

Our work focuses exclusively on structural polarization measures, which quantify polarization based solely on the network's topology.
This scope excludes methods that are non-structural or hybrid, meaning that they rely on node attributes, content, or auxiliary metadata.
For example, the Biased Random Walk (BRW)~\cite{emamgholizadeh2020framework} uses content-derived node embeddings to initialize and dissipate the random walk's \emph{energy}, and the Relative Closeness Controversy (RCC)~\cite{mendoza2020gene} applies NLP techniques to infer user bias towards named entities, producing an entity-conditioned graph rather than a purely structural one.

Our setting is also distinct from the literature on segregation, which addresses a related but conceptually different phenomenon.
Methods such as the Spectral Segregation Index (SSI)~\cite{echenique2007measure} quantify the isolation of a single group and are grounded in models of within-group social interaction (e.g., ghettos).
In contrast, \rwcpp targets information mixing between opposing poles.

Finally, we distinguish our work from research on polarization mitigation and network augmentation, which aims to modify the network to reduce its structural polarization.
For example, the work by Adriaens et al.~\cite{adriaens2023minimizing} focuses on minimizing hitting times across groups by adding shortcut edges. 
In contrast, our objective is strictly to develop a principled and robust measure of structural polarization, not to prescribe modifications.

For a comprehensive review and categorization of the other approaches proposed to quantify and mitigate network polarization, the interested reader may refer to the review by Interian et al.~\cite{interian2023network}.

\section{Limitations of Assortativity as a Polarization Measure}

Assortative mixing is known to influence network topology, and our experimental evaluation showed that it is also correlated with polarization (see \Cref{fig:roc_salloum} and \Cref{fig:roc_garimella}).
However, we remark that color assortativity does not have all the properties that we would desire from a polarization measure. For example, it is not zero on a bichromatic clique, even in the case when there are equal numbers of nodes of the two colors, rather it takes a slightly negative value. Being exactly zero on a clique, no matter the partition sizes, seems to us a fundamental requirement for a polarization measure, thus the use of assortativity to measure polarization is on shaky grounds, at least from a statistical principles point of view. Additionally, existing studies additionally indicate that assortativity alone is insufficient to explain the overall level of polarization of a network.
It may return the same values even when opinions are distributed in radically different mesoscale structures, such as communities~\cite{peel2018multiscale}, because its significance relies on most individuals' assortativity being close to the mean.

For example, in a study of collaboration networks, Leifeld~\cite{leifeld2018polarization} found that assortative mixing could indicate polarization, but interventions targeting it had little effect on the macroscopic polarization of the network. 
This resilience arises from the complex interplay of other microscopic factors (e.g., geographic proximity and topical similarity) beyond local mixing patterns.

Similarly, Hohmann et al.~\cite{hohmann2023quantifying} showed that assortativity is overly sensitive to initial structural changes when separation is weak, but loses sensitivity as structural separation becomes strong.
As a consequence, it struggles to distinguish weak communities from strong ones, making it insufficient for capturing the deep, global structural division characteristic of highly polarized networks.
In contrast, in such cases, RWC provides a more robust measure of structural division.

\section{Computational Complexity}

To obtain the stationary distributions of the RWRs used to instantiate $\procfun{v}$, we evaluated Personalized PageRank (PPR) from every vertex $v \in V$. Using the standard power-iteration solver, one PPR computation requires $\mathcal{O}(mk)$, where $m$ is the number of edges and $k$ is the number of iterations required for convergence (in our experiments, we set $k = 100000$). 
Thus, the exact computation of \rwcpp has total time complexity $\mathcal{O}(nmk)$. 
Our sampling-based approximation improves this by selecting only $s$ seed nodes and computing PPR from them. This reduces the complexity to $\mathcal{O}(smk)$.

\section{Additional Datasets}

\textsc{garimella}~\citep{garimella2016quantifying} contains $10$ controversial and $10$ non-controversial Twitter retweet networks, constructed from Tweets collected from Feb 27 to Jun 15, 2015, using sets of related hashtags, each anchored by one manually selected hashtag.
\textsc{conover}~\citep{conover2011political} consists of two networks constructed from tweets collected between Sep 14 and Nov 1, 2010, ahead of the U.S. congressional midterm elections. Starting from the two popular political hashtags \#p2 (“Progressives 2.0”) and \#tcot (“Top Conservatives on Twitter”), a set of 66 related hashtags was identified and used to create a retweet and a mention network.
For node labels, in \textsc{conover} we use the node labels provided by the authors, and for \textsc{garimella}, we generate node labels using the Kernighan--Lin (KLIN)~\citep{kernighan1970efficient} and METIS~\citep{karypis1997metis} partitioning algorithms.

\section{Additional Experiments}

\begin{figure}[!ht]
    \centering
    \includegraphics[width=\linewidth]{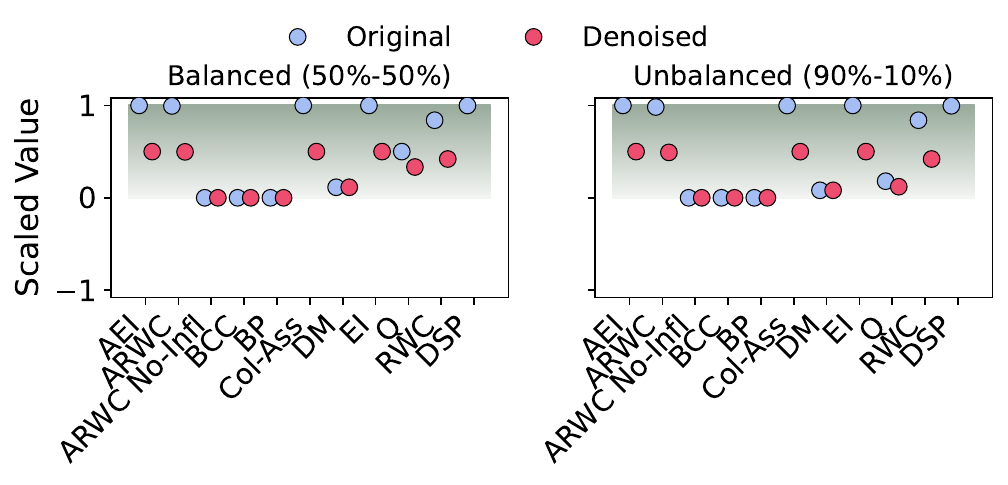}
    \caption{Polarization in a bi-colored half-split cycle with $5000$ vertices and different partition sizes: 50\% blue–50\% red (left), and 90\% red–10\% blue (right). We show the rescaled values and the rescaled denoised values computed using the $1k$-series~\citep{salloum2022separating}. The gradient area indicates the desired scores. }
    \label{fig:arwc-half-split}
\end{figure}

\spara{More on the ``no-influencers'' fix.}
As discussed in \Cref{sec:no-inf-fix}, excluding influencers from the restart set offers a simple way to remove the bias that \rwc exhibits on random graphs. Indeed, \Cref{fig:er-graphs-uneven} shows that, under this modification, the average \rwc score in random networks drawn from \gnml correctly converges to zero. However, this improvement does not generalize to more structured graph families. In particular, for the bi-colored half-split cycle with $5000$ vertices (\Cref{fig:arwc-half-split}), the adjusted score—ARWC No-Infl—collapses to~0, even though the expected value for this highly polarized topology is close to~1 (results for RWC No-Infl are similar). In other words, while the fix addresses the issue in random graphs, it simultaneously degrades performance on networks where strong polarization should be clearly detectable.

\begin{figure}[!ht]
    \centering
    \includegraphics[width=.9\linewidth]{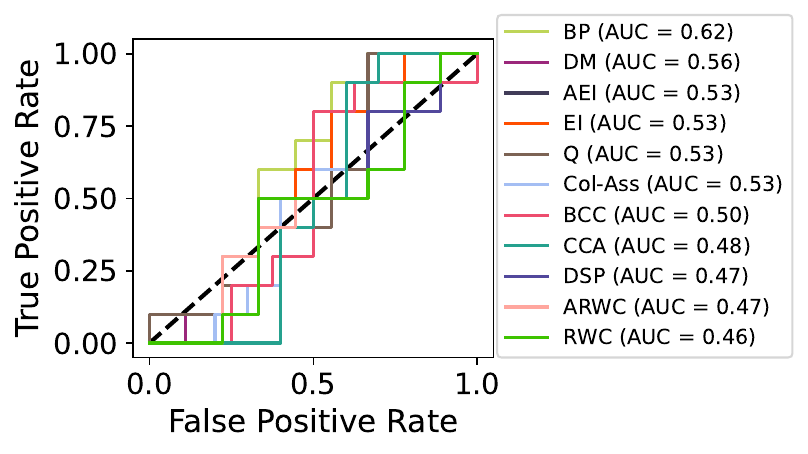}
    \caption{ROC curves and AUC values for \textsc{garimella}.}
    \label{fig:roc_garimella}
\end{figure}

\spara{More on Classification Performance.}
\Cref{fig:roc_garimella} shows that the performance of \rwcpp is comparable to that of \rwc.

\begin{figure}[!ht]
    \centering
    \includegraphics[width=0.49\linewidth]{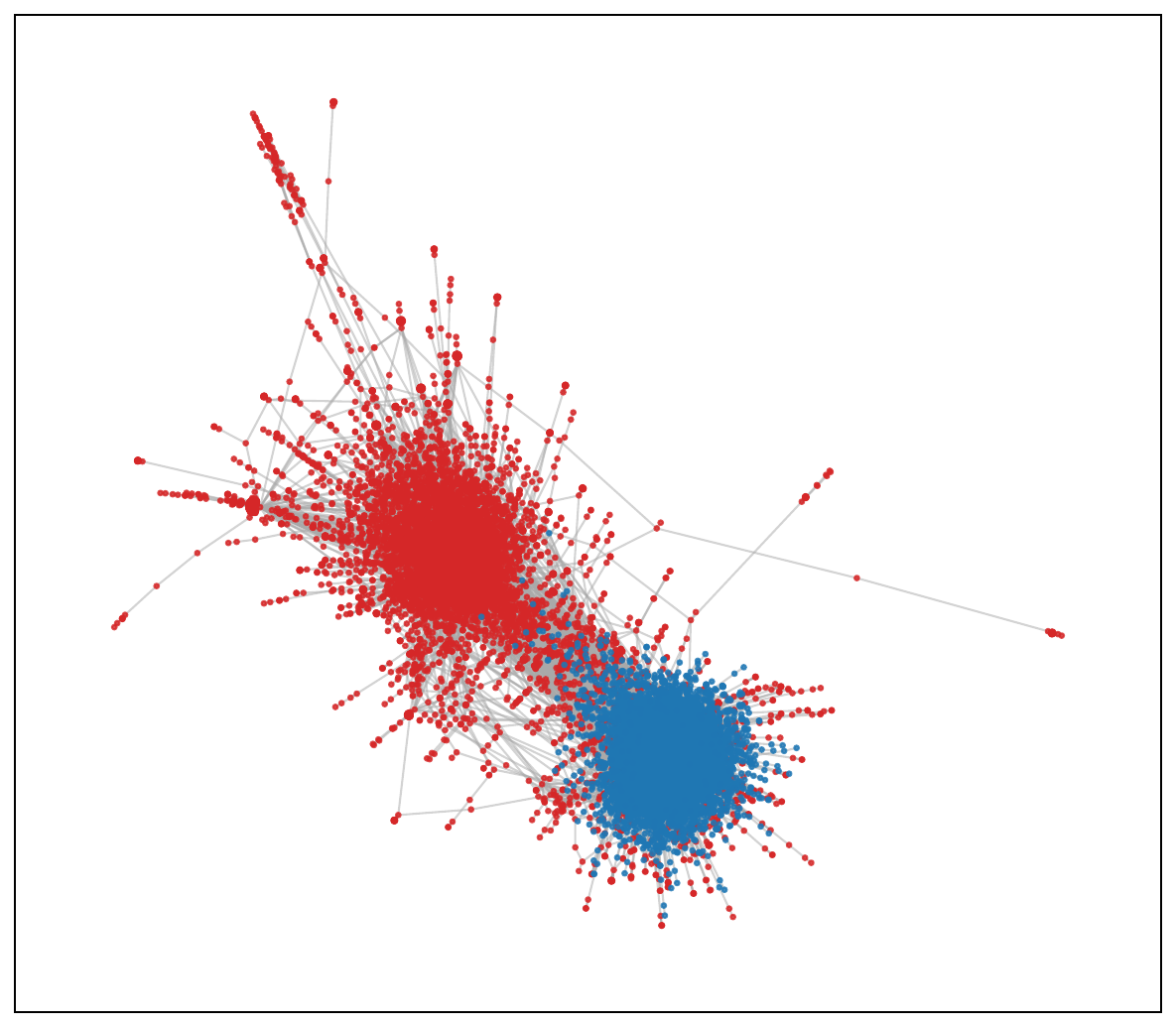}
    \includegraphics[width=0.49\linewidth]{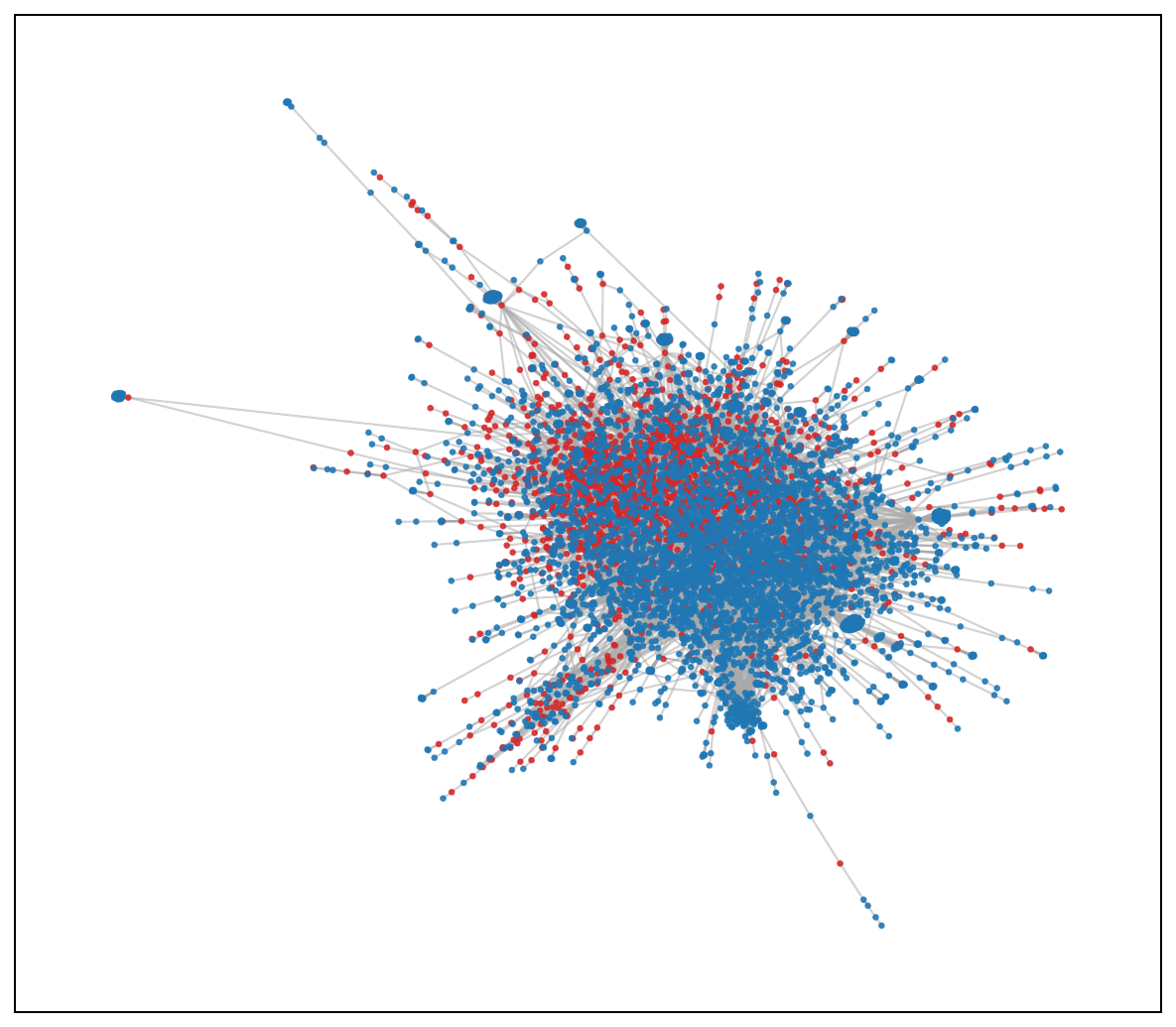}
    \caption{\textsc{retweet} (left) and \textsc{mention} (right) networks.}
    \label{fig:conover_nets}
\end{figure}

Similar results can be observed for the \textsc{conover} networks (\Cref{tbl:exp_conover}): higher values are recorded for the polarized \textsc{retweet} network (\textsc{R}), whereas values close to the minimum are recorded for the non-polarized \textsc{mention} (\textsc{M}) network (see \Cref{fig:conover_nets}).

\begin{table}[!ht]
    \caption{Polarization scores for the \textsc{conover} networks.}
    \label{tbl:exp_conover}
    \resizebox{\linewidth}{!}{
    \centering
    \begin{tabular}{lrrrrrrrrrrr}
    & \multicolumn{11}{c}{\textbf{Metric}} \\
    \cline{2-12}
    & AEI & ARWC & BCC & BP & CCA & Col-Ass & DM & EI & Q & \rwc & \rwcpp \\
    \midrule
    \textsc{M} & 0.306 & 0.150 & 0.609 & -0.030 & 0.451 & 0.227 & 0.643 & 0.315 & 0.101 & 0.101 & 0.057 \\
    \textsc{R} & 0.959 & 0.837 & 0.832 & 0.257 & 1.385 & 0.954 & 0.768 & 0.954 & 0.475 & 0.869 & 0.410 \\
    \bottomrule
    \end{tabular}
    }
\end{table}

\begin{figure}[!ht]
    \centering
    \includegraphics[width=.9\linewidth]{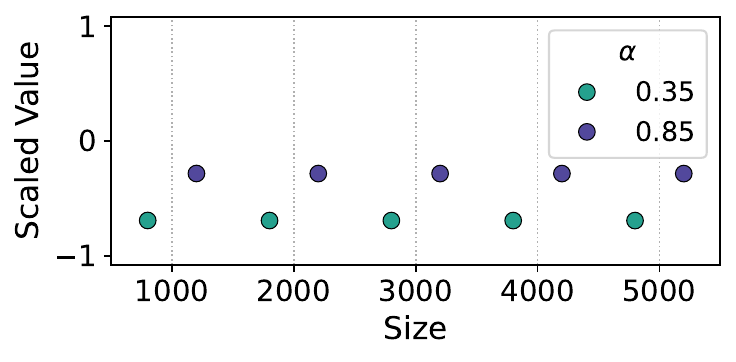}
    \caption{\rwcpp in bi-colored alternating cycles with various numbers of vertices, 50\% red and 50\% blue. We show the rescaled values for different values of $\alpha$.}
    \label{fig:vary_alpha}
\end{figure}

\spara{Impact of $\alpha$ on \rwcpp.}
\Cref{fig:vary_alpha} displays the \rwcpp scores computed on bi-colored alternating cycles with various numbers of vertices, evenly split between red and blue, using different values of the parameter $\alpha$ that controls the restart probability in the PPR computation. 
Lower values of $\alpha$ correspond to more frequent restarts, making the PPR vector more influenced by the immediate neighbors of the starting node. Higher values of $\alpha$ emphasize more central nodes.
We observe that \rwcpp is sensitive to the choice of $\alpha$: as $\alpha$ decreases, the \rwcpp score also decreases. This behavior is expected, as each node's immediate neighbors belong to the opposite community. 
When the diffusion process is more strongly influenced by neighbors, more probability mass flows across community boundaries, leading to a lower polarization score.

\begin{figure}[!ht]
    \centering
    \includegraphics[width=\linewidth]{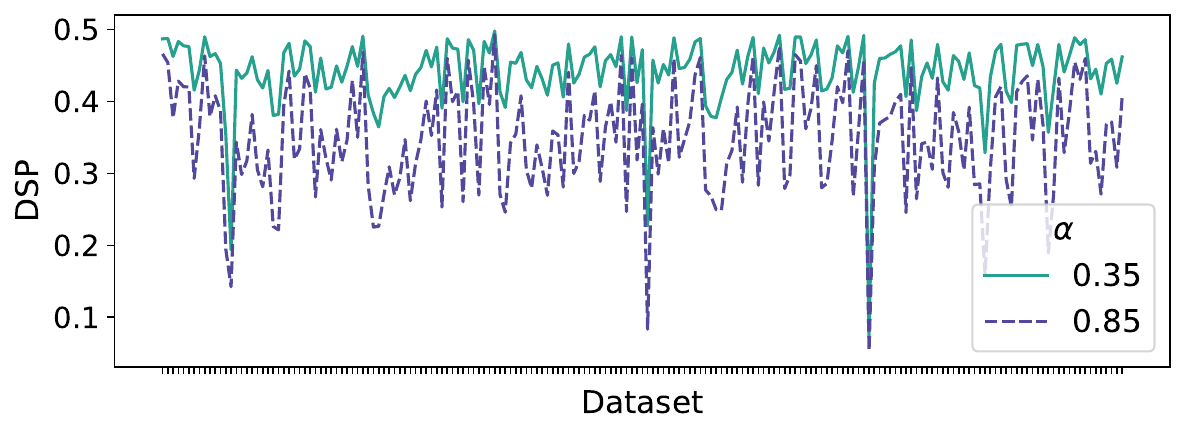}
    \caption{\rwcpp for \textsc{salloum}, varying the value of $\alpha$.}
    \label{fig:salloum_alpha}
\end{figure}

In addition, \Cref{fig:salloum_alpha} shows the value of \rwcpp for the \textsc{salloum} networks, for two values of $\alpha$.
Although the polarization scores are slightly lower for larger values of $\alpha$, the classification performance of \rwcpp remains stable ($0.68$ using $\alpha=0.85$ and $0.64$ using $\alpha=0.35$).

\begin{figure}[!ht]
    \centering
    \includegraphics[width=\linewidth]{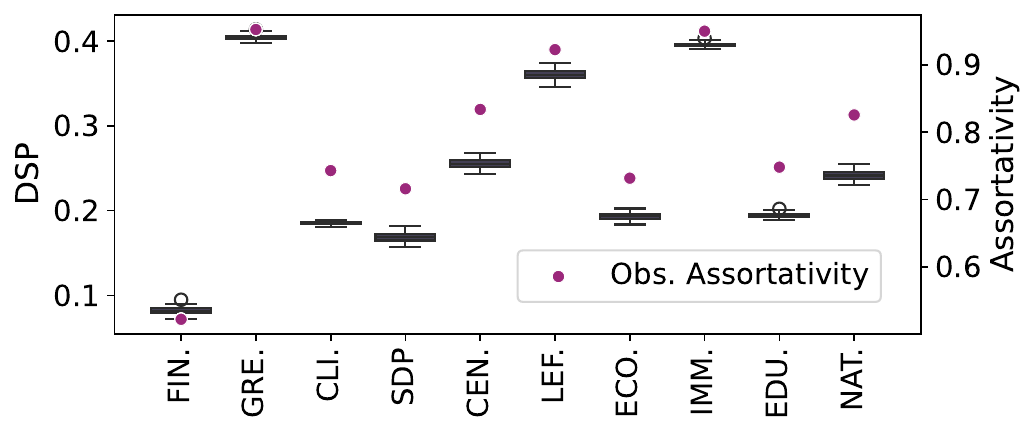}
    \caption{\textsc{salloum}: \rwcpp scores in the samples generated using Polaris and assortativity of the observed datasets.}
    \label{fig:polaris_exp_ax}
\end{figure}

\spara{More on Polarization and Assortativity.}
\Cref{fig:polaris_exp_ax} shows the distribution of the \rwcpp values in $100$ graphs sampled using Polaris, together with the color assortativity of the observed datasets. We report results for a subset of datasets for visualization purposes. This chart helps understand whether higher assortativity tends to correspond to higher polarization scores. We find that this relationship generally holds across datasets.

